%% file: Main.tex
\documentclass[review]{elsarticle}
\journal{arXiv}
\pdfoutput=1
\usepackage{graphicx,multirow,nomencl,framed,array,ragged2e}
\usepackage{amssymb}
\usepackage{amsmath}
\usepackage{graphicx}
\usepackage{algorithm}
\usepackage{array}
\usepackage[noend]{algpseudocode}
\usepackage{amsfonts, amsmath, amssymb}
\usepackage{stmaryrd}
\usepackage{booktabs,siunitx}
\usepackage[svgnames,table]{xcolor}
\usepackage[tableposition=above]{caption}
\usepackage{pifont}
\usepackage{bm}
\usepackage{cancel}
\usepackage{caption}
\usepackage{color}
\usepackage{enumerate}
\usepackage{float}
\usepackage{hyperref}
\usepackage[english]{babel}
\usepackage[margin=1in]{geometry}
\usepackage{mathtools}
\usepackage{setspace}
\usepackage{subfigure}
\usepackage{cancel}
\usepackage{lineno}

\begin{document}

\begin{frontmatter}
	
\title{Pulmonary drug delivery and retention: a computational study to identify plausible parameters based on a coupled airway-mucus flow model}
\author[JU1]{Aranyak Chakravarty}
\author[IITM2]{Mahesh V. Panchagnula}
\author[SVIMS]{Alladi Mohan}
\author[Northwestern1]{Neelesh A. Patankar\corref{mycorrespondingauthor}}
\ead{n-patankar@northwestern.edu}

\address[JU1]{School of Nuclear Studies \& Application, Jadavpur University, Kolkata, India}
\address[IITM2]{Department of Applied Mechanics, Indian Institute of Technology Madras, Chennai, India}
\address[SVIMS]{Department of Medicine, Sri Venkateshwara Institute of Medical Science, Tirupati, India}
\address[Northwestern1]{Department of Mechanical Engineering, Northwestern University, Evanston, IL, USA}
\cortext[mycorrespondingauthor]{Corresponding author}

\begin{abstract}
Pulmonary drug delivery systems rely on inhalation of drug-laden aerosols produced from aerosol generators such as inhalers, nebulizers etc. On deposition, the drug molecules diffuse in the mucus layer and are also subjected to mucociliary advection which transports the drugs away from the initial deposition site. The availability of the drug at a particular region of the lung is, thus, determined by a balance between these two phenomena. A mathematical analysis of drug deposition and retention in the lungs is developed through a coupled mathematical model of aerosol transport in air as well as drug molecule transport in the mucus layer. The mathematical model is solved computationally to identify suitable conditions for the transport of drug-laden aerosols to the deep lungs. This study identifies the conditions conducive for delivering drugs to the deep lungs which is crucial for achieving systemic drug delivery. The effect of different parameters on drug retention is also characterized for various regions of the lungs, which is important in determining the availability of the inhaled drugs at a target location. Our analysis confirms that drug delivery efficacy remains highest for aerosols in the size range of 1-5 $\mu$m. Moreover, it is observed that amount of drugs deposited in the deep lung increases by a factor of 2 when the breathing time period is doubled, with respect to normal breathing, suggesting breath control as a means to increase the efficacy of drug delivery to the deep lung. A higher efficacy also reduces the drug load required to be inhaled to produce the same health effects and hence, can help in minimizing the side effects of a drug.
\end{abstract}


\end{frontmatter}

\section*{Introduction}
The lung is one of the most exposed organs of the human body \cite{west2012respiratory}. The dichotomous branching structure of the lung - starting from the trachea and culminating in the alveolar sacs - provides a mechanism by which air from the surrounding atmosphere is drawn into the lungs during inhalation and expired out during exhalation. Pulmonary drug delivery systems take advantage of the respiration process to deliver drug molecules to the lung through inhalation. The drug molecules may be in the form of dry powders or liquid aerosols, and are administered in a non-invasive manner with the help of aerosol generators such as inhalers, nebulisers etc. \cite{kooij2019size,mohandas2021overview}. Once inhaled, the powdered/aerosolised drugs are transported along the respiratory tract where they deposit depending on their physio-chemical properties as well as breathing characteristics and physiological conditions. Thus, drugs can be delivered locally to a targeted region of the lung for treatment of respiratory diseases, such as asthma or COPD \cite{mohandas2021overview}. Such targeted delivery can potentially lead to smaller overall drug dose and reduced side effects. Systemic drug
delivery can also be achieved by targeting delivery to the alveolar region of the lung where the drugs can be easily absorbed into the systemic blood circulation through the thin blood-gas barrier and the large alveolar surface area \cite{west2012respiratory}.

The transport of the inhaled aerosols within the respiratory tract is governed by the combined effects of unsteady convective air flow, gravitational settling, and aerosol diffusion in air \cite{hofmann2011modelling}. At the same time, the inhaled aerosols are deposited primarily due to diffusion, sedimentation, and inertial impaction \cite{mittal_jfm,guha2008transport,hofmann2011modelling}, which depend significantly on aerosol properties and other physiological parameters \cite{hofmann2011modelling}. It has been observed that a major portion of the inhaled aerosols are deposited in the naso-pharyngeal region \cite{hofmann2011modelling}. Deposition may also take place in other regions of the respiratory tract before the inhaled aerosol particles reach the target region. This effectively reduces the actual dose reaching the target region of the lung. For example, aerosols larger than 10 $\mu$m have been observed to be completely deposited in the upper respiratory tract and do not reach the alveolar region at all \cite{devi2016designing, choi2007mathematical}. The physio-chemical properties (size, shape, morphology, chemical composition etc.) of the inhaled aerosol must, as such, be tailored to facilitate drug delivery to the target region depending on breathing characteristics and other physiological conditions. 

Different techniques (Eulerian, Lagrangian and combinations thereof) have been used to computationally model aerosol transport and deposition \cite{hofmann2011modelling} in specific regions of the respiratory tract\cite{chakravarty2019aerosol,fishler2015particle,koullapis2018efficient} as well as the whole lung. Here, \textit{whole lung} models consider the lungs to be a network of interconnected branching channels with varying dimensions based on lung morphometry. The computational model used in the present analysis is based on one such \textit{whole lung} model \cite{taulbee1975theory,devi2016designing} - based on a Weibel \cite{weibel1963morphometry} lung geometry with appropriate modifications.

The inhaled aerosols, containing the drug molecules, are deposited in the respiratory mucus \cite{mauroy2011toward, karamaoun2018new}. The mucus layer lines the inner surface of the respiratory tract and prevents the deposited materials from coming in direct contact with the epithelial cells (which lie underneath the mucus lining) and the capillaries (which remain beyond the epithelium) \cite{mauroy2011toward}. The respiratory mucus, therefore, acts as a barrier to drug absorption. In addition, the epithelial cells are also lined with cilia which beat metachronously within the periciliary layer \cite{karamaoun2018new} transporting the mucus, and the deposited materials, from the distal airways towards the pharyngeal region. Mucociliary clearance, as such, further prevents effective absorption of the deposited drug molecules. It is, therefore, essential to consider mucociliary transport while studying drug delivery in the lungs. However, mathematical models published in the literature have not accounted for mucociliary transport while investigating pulmonary drug delivery. 

Thus, in order to computationally explore the pulmonary drug delivery mechanism, one needs a mathematical model that takes into account aerosol transport (in airways) and drug molecule transport (in mucus), since these transport processes occur simultaneously within the lung. Such a model is being reported for the first time. This article reports such a model within the framework of a Weibel model of the human lung. The primary goal is to use this mathematical model to identify situations that can lead to the transport of aerosols, containing the drug molecules, from the pharyngeal region to the deep lungs. The model is also used to determine the conditions that promote retention of the deposited drug molecules in the lungs and thereby, increases the bioavailability of the drugs.

Although the mathematical model has been used here to specifically study drug delivery to the lungs, the same model can be utilised to study other similar physical processes involving exposure of the lungs to foreign particles such as pollutant (smoke, dust etc.) and pathogen (virus, bacteria etc.) deposition and clearance from the lungs. 

\section*{Idealisation of the lung geometry}
\label{sec:lunggeometry}
\begin{figure}[!ht]
    \centering
    \includegraphics[scale=0.8]{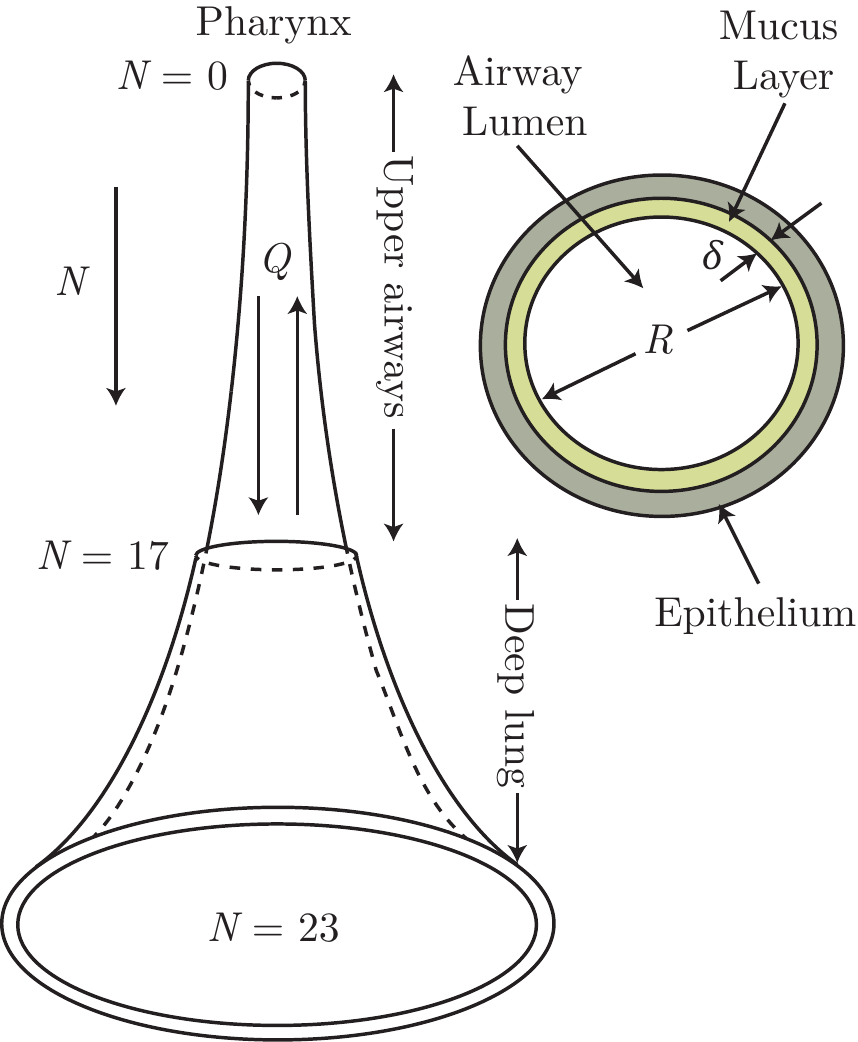}
    \caption{Schematic illustration of the one-dimensional \textit{trumpet} model that is used in the present analysis to approximate the dichotomous network structure of a human lungs. A cross-sectional view of a single airway branch is also shown to illustrate the arrangement of the airway lumen, mucus layer, and epithelial lining in model.}
    \label{fig:schematic}
\end{figure}

The physiological dichotomous branching network of human lungs is approximated in this work by a one-dimensional \textit{trumpet} model (Fig. \ref{fig:schematic}). While this model cannot account for the effects of heterogeneity in the lungs, it is still a tractable model for the whole lungs in order to capture key trends.

The airway is modeled as a continuous one-dimensional channel of variable cross-sectional area, where the length is divided into 24 generations ($N=0-23$; $N$ is the generation number), based on morphometric data of a human lungs \cite{weibel1963morphometry}. For a dichotomous tree, the number of bronchioles in each generation is $2^N$, while the length ($L$) and the total cross-sectional area ($A$) at each generation is calculated using a power-law function as
\begin{equation}
    L(N) = L_0\alpha^N,A(N) = A_0(2\beta)^N,
\end{equation}
where $L_0$ and $A_0$ are the length and cross-sectional area at $N=0$, respectively (see Table S1 for magnitudes). The length-change ($\alpha$) and area-change ($\beta$) factors are selected (Table S1) such that the computed length and area at each generation matches Weibel's morphometric data \cite{weibel1963morphometry}. Although $N$ is an integer, it is treated as a continuous variable in all transport equations for computational convenience. The airway length ($x$), in terms of the lung generation number $N$, is given by 
\begin{equation}
    x(N) = \frac{L_0(1-\alpha^{N+1})}{1-\alpha}.
\end{equation}

Alveolation of the airways is considered $N=17$ onwards, consistent with human lungs \cite{weibel1963morphometry}, by adding area in the relevant generations (see Table S2). 

The modeled system of airways and alveoli is also assumed to be lined by a thin mucus layer separating the airway lumen from the epithelium. Mucociliary transport is accounted for by assuming a convective motion of the mucus layer from the deeper generations towards the $0^{th}$ generation. The thickness ($\delta$), the total cross-sectional area of the mucus layer ($A_m$), and the convective mucus velocity ($V_m$) at different lung generations are estimated as

\vspace{-10pt}
\begin{equation}
\begin{split}
    \delta(N) = \delta_{0}\zeta^N,\ &A_m(N) = A_{m,0}(2\sqrt{\beta}\zeta)^N,\\
    V_m(N) & = V_{m,0}\varepsilon^N \text{, for N}<18,\\
    & = 0 \text{, for N}\geq18,
\end{split}
\label{eq:mucus}
\end{equation}
where $\delta_{0}$, $A_{m,0}$, and $V_{m,0}$ are the mucus thickness, area, and velocity at $N=0$, respectively (see Table S1). The magnitudes of the change factors $\zeta$ and $\varepsilon$ (see Table S1) are chosen based on experimental data \cite{karamaoun2018new}. $V_m$ is zero beyond $N=18$ (Eq. \ref{eq:mucus}) due to the absence of appreciable mucociliary transport in the deep lungs \cite{mauroy2011toward}. $\delta$ and $V_m$ are also assumed to be temporally invariant in this analysis. \cite{karamaoun2018new}.

\section*{Mathematical model}
\label{sec:math_model}

\subsection*{Aerosol transport in airways}
\label{sec:aerosol transport eqn}

The one-dimensional transport equation for aerosols in the idealized lung geometry is 
\begin{equation}
    \frac{\partial(Ac_a)}{\partial t}+H\frac{\partial(Qc_a)}{\partial N} = H\frac{\partial }{\partial N}\Big(A D_aH\frac{\partial c_a}{\partial N}\Big)-L_Dc_a,
    \label{eq:aerosol_tr_1}
\end{equation}
where $c_{a}$, $Q$, and $D_{a}$ are aerosol concentration, volume flow rate of air during breathing, and aerosol diffusivity in air, respectively, and $H(N) = \dfrac{\partial N}{\partial x}$. The coefficient $L_{D}$ models aerosol deposition in the airway mucus. Eq. \ref{eq:aerosol_tr_1} assumes that the aerosols are monodispersed, do not coagulate, and do not affect the airflow in the lungs. Consistent with the focus of this study, it is assumed that the only source of aerosols is at the entrance to the $0^{th}$ generation, presumably from an aerosol generator. No additional aerosolization of the mucus or aerosol source are considered within the lungs. The inhaled aerosols are either deposited or washed out of the airways. Eq. \ref{eq:aerosol_tr_1} is reduced to a dimensionless form (Eq. \ref{eq:aerosol_transport_final}) using scalings defined in Eq. \ref{eq:aerosol_scaling} below (see \textit{Supplementary Materials})

\begin{equation}
\begin{split}
\centering
    \tau = &\frac{t}{T_b}, 
    \phi_a = \frac{c_a}{c_{a,0}},
    T_a = \frac{L_0 A_0}{|Q_{max}|},
    St_a = \frac{T_a}{T_b},\\
    &Pe_{a} = \frac{|Q_{max}|L_0}{A_0D_{a}},D_{a}=\dfrac{k_B T C_S}{3\pi\mu_{a}d_{a}}.
\end{split}
\label{eq:aerosol_scaling}
\end{equation}

\begin{equation}
\begin{aligned}
    Pe_{a}St_a(2\alpha\beta)^N &\frac{\partial( \phi_a)}{\partial \tau}=\frac{\partial }{\partial N}\Bigg[\Bigg(\Bigg(\frac{2\beta}{\alpha}\Bigg)^N \Bigg(\frac{1-\alpha}{\alpha (\text{ln} \alpha)}\Bigg)^2 \frac{\partial \phi_a}{\partial N}\Bigg)\\&+\Big(Pe_{a}q(t)\Big(\frac{1-\alpha}{\alpha \text{ ln}(\alpha)}\Big)\phi_a\Big)\Bigg]-L'_D\phi_a,
\end{aligned}
\label{eq:aerosol_transport_final}
\end{equation}
where $q(t)$ is a sinusoidal function accounting for airflow variation during breathing ($Q = Q_{max}q(t)$). $Pe_{a}$, $St_a$, $\phi_{a}$, and $\tau$ represent aerosol Peclet number, airway Strouhal number, dimensionless aerosol concentration, and dimensionless time, respectively. Note that $Pe_{a}$ refers to the aerosol Peclet number at $N=0$ only. As such, even if $Pe_{a}$ is extremely large, the local Peclet numbers at the higher generations can remain small. $T_a$ is the convective airflow timescale and $T_b$ is the breathing time period. $D_{a}$ is calculated using the Stokes-Einstein relation, where $k_B$, $T$, $C_S$, $\mu_{a}$, and $d_{a}$ are the Boltzmann constant, temperature, Cunningham slip correction factor, viscosity of air, and aerosol diameter, respectively \cite{chakravarty2019aerosol}. $L'_D$ is the dimensionless aerosol deposition coefficient which is determined using empirical models for various deposition mechanisms (see \textit{Supplemental Materials}).

\subsection*{Drug molecule transport in mucus}
\label{sec:particle transport eqn}

The one-dimensional transport equation for the deposited drug molecules in the mucus is formulated considering mucociliary transport and diffusion of the deposited drug molecules in the mucus. It is expressed as   
\begin{equation}
    \frac{\partial(A_m c_{d})}{\partial t}+H\frac{\partial(Q_m c_{d})}{\partial N} = H\frac{\partial }{\partial N}\Big(A_m D_{d}H\frac{\partial c_{d}}{\partial N}\Big)+L_Dc_a\phi_l,
\label{eq:part_tr_1}
\end{equation}
where $c_{d}$, $Q_m$, and $D_{d}$ are drug concentration in the mucus, volume flow rate of mucociliary transport, and drug molecule diffusivity in the mucus, respectively. $\phi_l$ is the drug load in the droplets, defined as the quantity of drug molecules contained per unit quantity of droplets. The term $L_{D}c_{a}\phi_l$ takes into account the drug molecules being introduced into the mucus due to aerosol deposition. Further absorption of the deposited drugs across the epithelium into the blood stream is not considered presently. Eq. \ref{eq:part_tr_1} is converted to a dimensionless form (Eq. \ref{eq:part_tr_final}) using scalings defined in Eq. \ref{eq:particle_scaling} below (see \textit{Supplementary Materials})
\begin{equation}
\begin{aligned}
    \tau =& \frac{t}{T_b}, 
    \phi_{d} = \frac{c_d}{c_{d,0}},
    c_{d,0} = \phi_l c_{d,0} \dfrac{A_0}{A_{m,0}},
    T_m = \frac{L_0}{|V_{m,0}|},\\
    &St_m = \frac{T_m}{T_b},
    Pe_{d} = \frac{|V_{m,0}|L_0}{D_d}, D_d = \dfrac{k_B T}{3\pi\mu_m d_d}.
\end{aligned}
\label{eq:particle_scaling}
\end{equation}

\begin{equation}
\begin{aligned}
    Pe_{d}(2\alpha\zeta\sqrt{\beta})^N &St_m \frac{\partial \phi_d}{\partial \tau}=\\&
    \frac{\partial }{\partial N}\Bigg[\Bigg(\Big(\frac{2\zeta\sqrt{\beta}}{\alpha}\Big)^N \Big(\frac{1-\alpha}{\alpha \text{ ln}(\alpha)}\Big)^2 \frac{\partial \phi_d}{\partial N}\Bigg)\\&-\Bigg(Pe_{d}(2\varepsilon\zeta\sqrt{\beta})^N \phi_d)\Bigg)\Bigg]+\Big(L'_{D}\frac{D_a}{D_d}\phi_a\Big),
\end{aligned}
\label{eq:part_tr_final}
\end{equation}
where $\phi_d$, $Pe_d$, and $St_m$ are the dimensionless drug concentration, drug Peclet number, and mucus layer Strouhal number, respectively. Also note that $Pe_d$ refers to the virus Peclet number at $N=0$ only. $T_m$ denotes the time-scale for mucociliary transport. $D_d$ is estimated using the Stokes-Einstein relation, where $\mu_{m}$ and $d_{d}$ are the viscosity of the mucus and the drug molecule diameter, respectively. The last term on the right hand side of Eq. \ref{eq:part_tr_final} is the dimensionless drug source due to aerosol deposition.

\subsection*{Initial and boundary conditions}

The lungs are assumed to be initially devoid of aerosols and drugs, i.e., $\phi_{a}|_{\tau=0} = \phi_{d}|_{\tau=0} = 0$ at all generations. It is also assumed that $N=0$ of the lungs is exposed to drug-laden aerosols, presumably from an aerosol generator, for a specific exposure duration ($\tau_{exp}$). The aerosols are breathed in during inhalation (Eq. \ref{eq:BC1}) and washed out during exhalation (Eq. \ref{eq:BC2}). In contrast, the drugs are always assumed to be washed out of $N=0$, along with the mucus, irrespective of inhalation/exhalation (Eq. \ref{eq:BC3}). At the distal end of the lungs ($N=23$), the total advection-diffusion flux of both aerosols and drugs is assumed to be zero (Eq. \ref{eq:BC4}). Mathematically, these conditions are expressed as follows 
\begin{equation}
    \begin{split}
        \phi_{a}\big|_{N=0} &= 1, \tau\leq\tau_{exp}, \\
    &= 0, \tau>\tau_{ exp},
    \end{split}
    \label{eq:BC1}
\end{equation}

\begin{equation}
    \dfrac{\partial (F_{a})}{\partial N}\Big|_{N=0}=0, \tau>0,
    \label{eq:BC2}
\end{equation}

\begin{equation}
    \dfrac{\partial (F_{d})}{\partial N}\Big|_{N=0}=0, \tau>0,
    \label{eq:BC3}
\end{equation}

\begin{equation}
    F_{a}\big|_{N=23} = F_{d}\big|_{N=23}=0, \tau>0,
    \label{eq:BC4}
\end{equation}
where $F_{a}$ and $F_{d}$ are the total advection-diffusion flux in the aerosol transport (Eq. \ref{eq:aerosol_transport_final}) and drugs transport equation (Eq. \ref{eq:part_tr_final}), respectively (see \textit{Supplementary Materials}). Detailed derivation of the mathematical model and its validation (\textit{Fig. S2}) are provided in \textit{Supplementary Materials}.

\section*{Results and discussion}
\label{sec:results}
\begin{figure*}[!ht]
    \centering
    \includegraphics[scale=0.85]{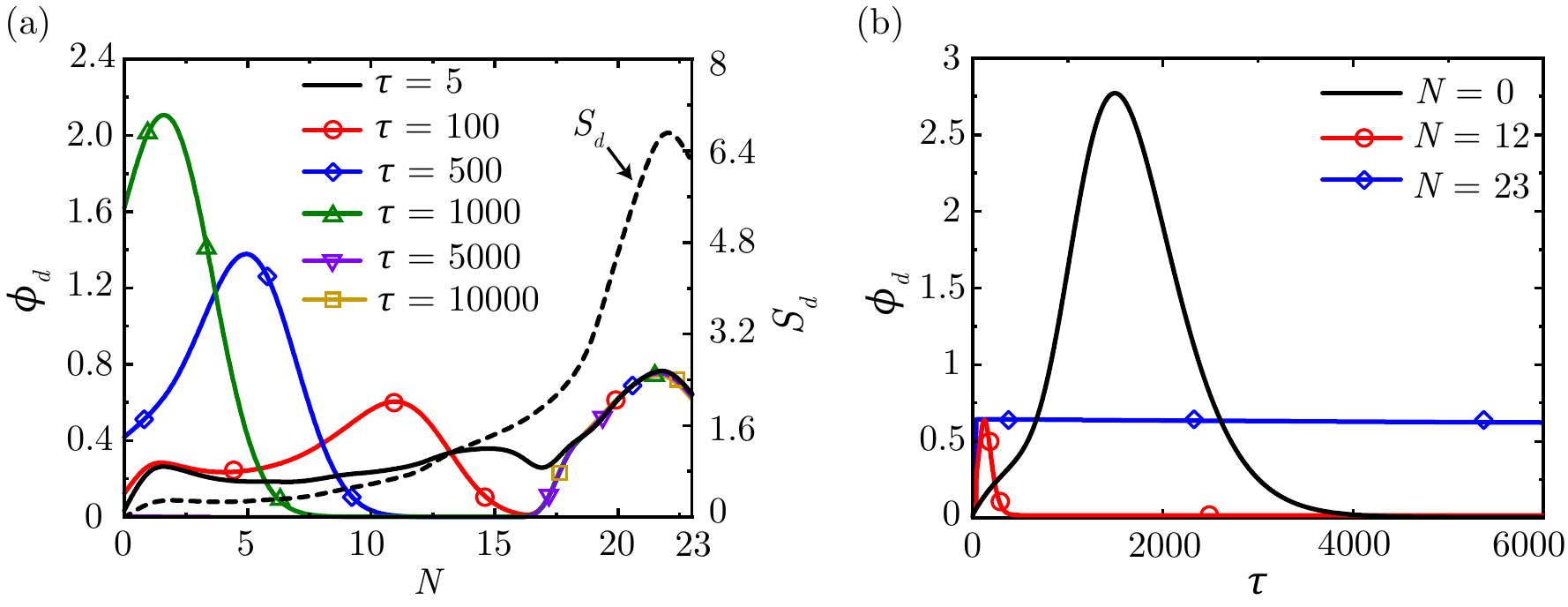}
    \caption{(a) Aerosol deposition ($S_d$ $=\int \int L'_{D}\phi_d d\forall d\tau$) within the lungs at the end of exposure and drug concentration ($\phi_{d}$) within the lungs at different time instances ($\tau$) (b) Temporal change in $\phi_{d}$ at $N=0,12,23$. The results are shown for $Pe_{a} = 2.85 \times 10^{10}$, $St_a = 0.0095$, $Pe_{d} = 4.56 \times 10^{7}$, $St_m = 359.7122$, $\tau_{exp}=5$.}
    \label{fig:particle_transport}
\end{figure*}

Drug-laden aerosols are deposited in the respiratory mucus  primarily during inhalation. The deposited drug molecules diffuse in the mucus layer and are transported upstream (towards the mouth) via mucociliary advection. To obtain the key deposition and washout trends, simulations were done assuming that drug-laden aerosols are entering the lungs for five breaths, i.e., exposure time $\tau_{exp}=5$. Extrapolation to longer exposure times and its impact on drug retention will be discussed separately. It is seen that the (scaled) drug concentration in the mucus ($\phi_{d}$), at the end of the exposure duration ($\tau=5$), qualitatively follows aerosol deposition $S_d$ ($=\int \int L'_{D}\phi_a d\forall d\tau$; see Fig. \ref{fig:particle_transport}a). 
 
Drug molecules deposited in the conducting airways ($N<18$; $N$ represents the lung generation) is transported upstream towards the mouth ($N=0$).
This results in higher drug concentration $\phi_{d}$ in the upper airways (lower $N$) primarily due to smaller mucus volume. Eventually, the drugs are washed out of the lungs (see Fig \ref{fig:particle_transport}a). The temporal change in $\phi_{d}$ at the mouth (Fig. \ref{fig:particle_transport}b) also corroborates this conclusion. 

In contrast, drugs deposited in the deeper generations ($N\geq18$) are not subjected to mucociliary transport. Therefore, $\phi_{d}$ undergoes a gradual change due to weak diffusive transport. As such, drugs deposited in the deep lungs persist for a much longer time as compared to that deposited in the upper airways. This is also clearly evident from Fig. \ref{fig:particle_transport}.    

Deep lung (alveolar) deposition of the drugs is beneficial for systemic drug delivery primarily due to the thin mucus layer in the deep lung and the large surface area of the alveoli and the alveolated bronchioles in contact with the blood vessels. This enables the deposited drugs to come in close contact with the blood vessels and increases the probability of the drugs entering the blood stream, thereby ensuring systemic drug delivery. A longer residence time of the deposited drugs within the deep lungs further increases the probability of systemic drug delivery. Thus, it is important to understand the various effects that cause the drugs to deposit and persist in the deep lungs. This is discussed next. Physiologically relevant ranges are chosen for all parameters in this study (see \textit{Tables S1} and \textit{S3} in the \textit{Supplemental Material} for more details).   

\subsection*{Effect of aerosol size on drug deposition in the deep lungs}
\label{sec:aerosol_peclet}

\begin{figure*}[!ht]
    \centering
    \includegraphics[scale=0.85]{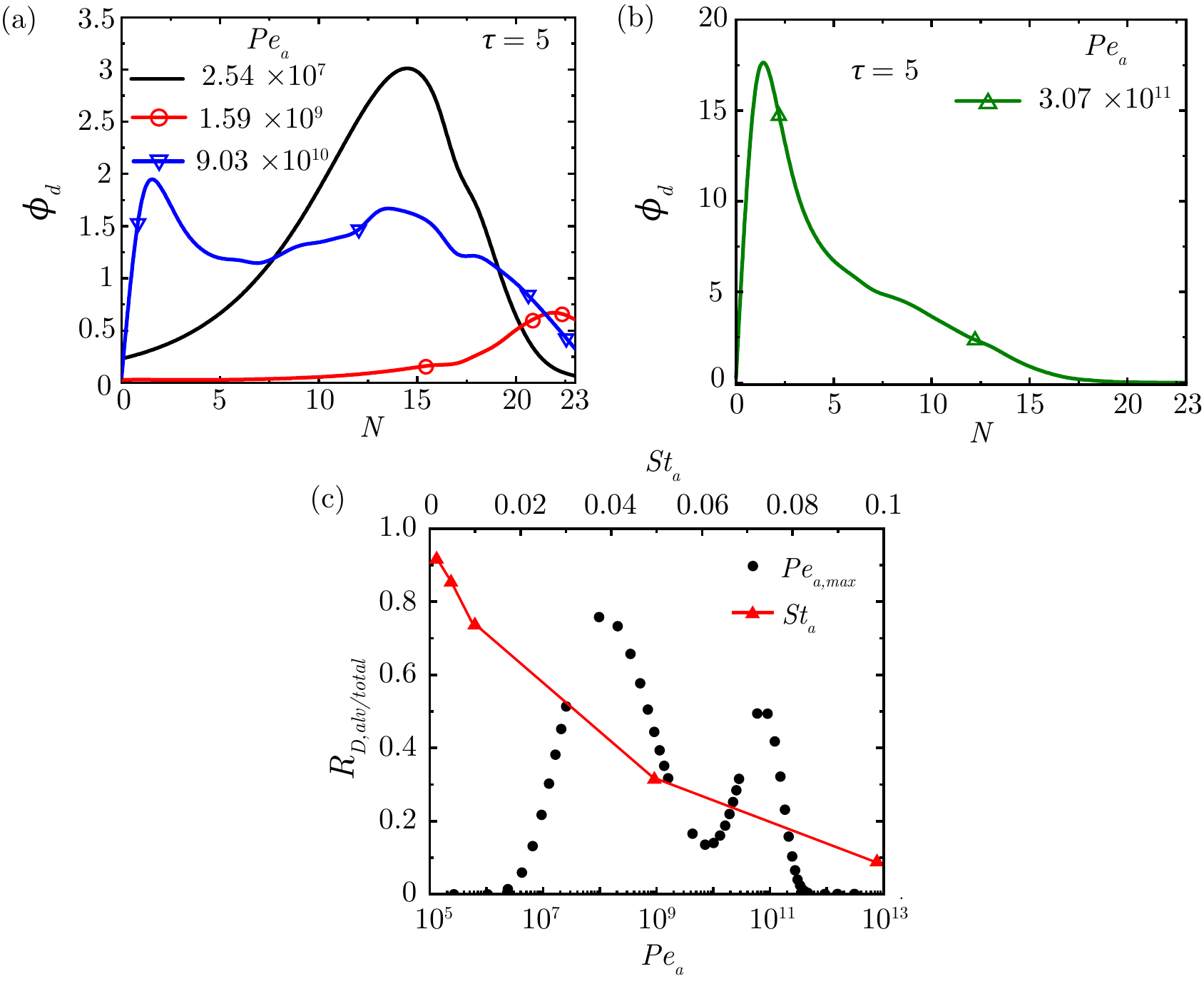}
    \caption{(a-b) $\phi_{d}$ within the lungs for different $Pe_{a}$ at the end of exposure ($St_a = 0.0095$, $Pe_{d} = 4.56 \times 10^{7}$, $St_m = 359.7122$, $\tau_{exp}=5$) (c) Change in fraction of droplets deposited in the deep lungs to that in the whole lungs ($R_{D,alv/tot}$) with variation in $Pe_{a}$ and $St_a$.}
    \label{fig:Pe_ae}
\end{figure*}

\begin{figure*}[!ht]
    \centering
    \includegraphics[scale=0.85]{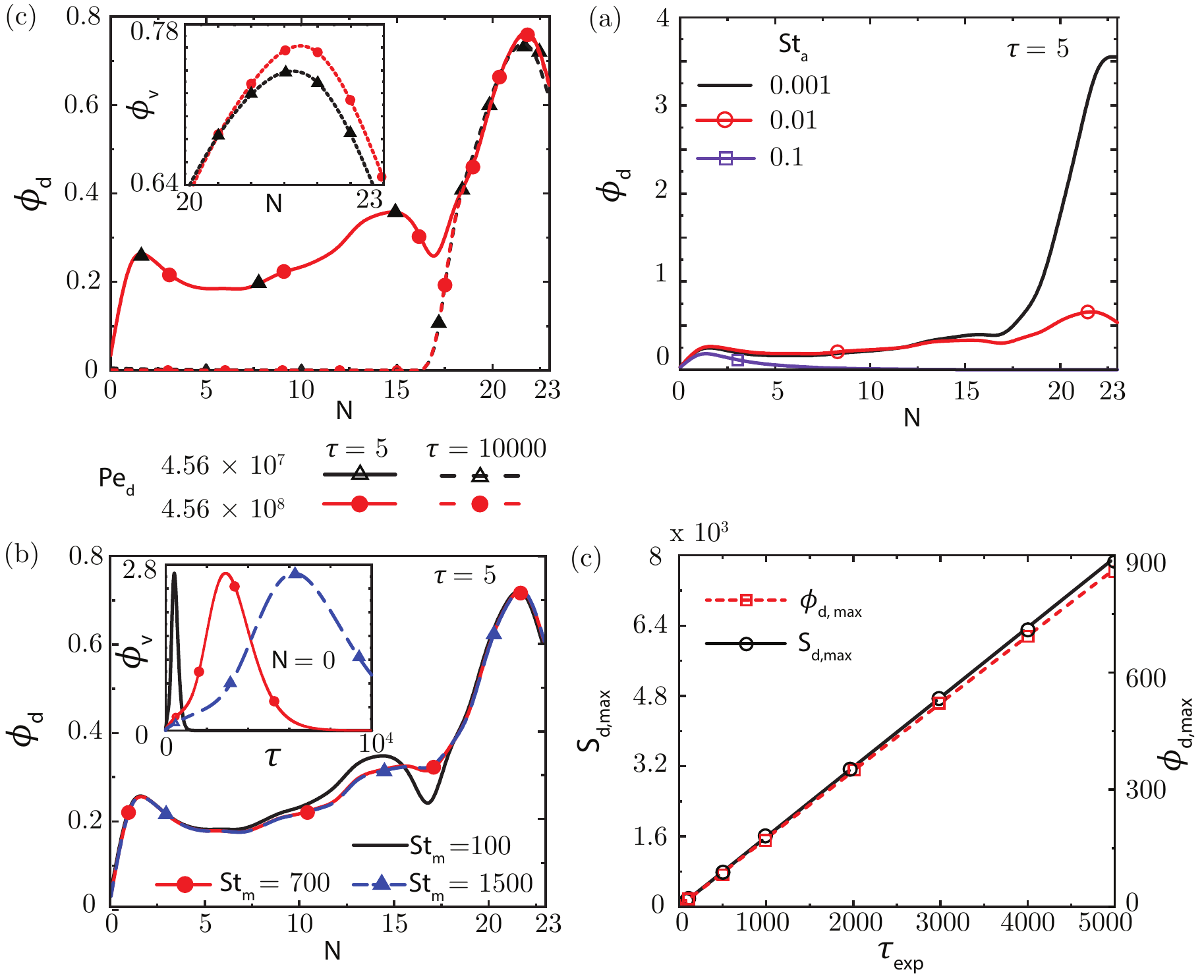}
    \caption{(a) $\phi_{d}$ within the lungs at the end of exposure ($\tau=5$) and at $\tau=10000$ for two different $Pe_{d}$ ($Pe_a = 2.85\times10^{10}$, $St_a = 0.0095$, $St_m = 359.7122$, $\tau_{exp}=5$). A zoomed view of $\phi_{d}$ in the deep lungs is shown as inset to adequately highlight the difference in $\phi_{d}$ for the two cases (b) $\phi_{d}$ within the lungs for different $St_a$ at the end of exposure to drug-laden aerosols ($Pe_{a} = 2.85 \times 10^{10}$, $Pe_{d} = 4.56 \times 10^{7}$, $St_m = 359.7122$, $\tau_{exp}=5$) (c) $\phi_{d}$ within the lungs at the end of exposure for various $St_m$ at $\tau=5$ ($Pe_{a} = 2.85 \times 10^{10}$, $Pe_{d} = 4.56 \times 10^{7}$, $St_a = 0.0095$, $\tau_{exp}=5$). The temporal change of $\phi_{d}$ at $N=0$ is shown as inset to highlight faster drug washout from the upper airways at smaller $St_m$. (d) Increase in aerosol deposition ($S_d$) and $\phi_{d}$ within the lungs with rise in exposure time ($Pe_{a} = 2.85 \times 10^{10}$, $Pe_{d} = 4.56 \times 10^{7}$, $St_a = 0.0095$, $St_m=359.7122$).}
    \label{fig:Pe_vi}
\end{figure*}

Aerosol Peclet number ($Pe_{a}$) is defined as the ratio of advective transport to diffusive transport of aerosols in air (see Eq. \ref{eq:aerosol_scaling}). Greater peak inspiratory flow rate will lead to larger $Pe_{a}$, which implies greater advective transport. Smaller aerosols exhibit greater diffusive transport leading to smaller $Pe_{a}$. Intuitively, one would expect the aerosols to reach deeper parts of the lungs at larger $Pe_{a}$ due to stronger advective transport in air. However, deposition trends are non-monotonic (see Fig. \ref{fig:Pe_ae}a-b). Specifically, deposition in the deep lungs increases up to $Pe_{a}=1.59 \times 10^{9}$ and then decreases. Additionally, the peak of $\phi_{d}$ is observed in lower generations ($N<18$) at both small and large values of $Pe_{a}$. This is because, at small $Pe_{a}$, the advection is not strong enough to carry the aerosols into the deep lungs, whereas at large $Pe_{a}$ the aerosols deposit in the upper airways due to the impaction mechanism (see Fig. \ref{fig:Pe_ae}b). Drug retention within the lungs, however, remains unaffected when $Pe_{a}$ is changed, since it affects neither mucociliary transport nor drug diffusivity in mucus (see \textit{Supplementary Material, Fig. S3} for more details).   

Fig. \ref{fig:Pe_ae}c shows the fraction of the drug-laden aerosols deposited in the deep lungs at different values of $Pe_{a}$. It is seen that deposition of the aerosols in the deep lungs occurs when $2.37 \times 10^{6}<Pe_{a}<3.07 \times 10^{11}$. This range translates to aerosol diameters of 10 $\mu$m to 0.003 $\mu$m for normal breathing in a healthy individual (tidal volume of $1000 \text{ ml}$ and $T_b=4 \text{ s}$). Within this range, deposition is comparatively less for $4.29 \times 10^9<Pe_{d}<1.6 \times 10^{10}$ (aerosols diameters $\sim 0.2 - 0.6 \text{ }\mu$m).

In summary, aerosols smaller than 10 $\mu$m diameter will tend to deposit in the deep lungs under normal breathing conditions. Typical aerosol sizes obtained from aerosol generators (inhalers, nebulizers etc.) are in the range of $0.1 - 100\text{ }\mu$m~\cite{mohandas2021overview}.

\subsection*{Effect of mucus advection and viscosity on drug retention}
\label{sec:particle_peclet}
Drug Peclet number ($Pe_{d}$) is the ratio of advective mucociliary transport and diffusive transport of the drug molecules in the mucus layer (see Eq. \ref{eq:particle_scaling}). An increase in $Pe_{d}$ indicates a larger contribution of mucociliary transport (or a smaller impact of diffusion) in the overall transport process and vice-versa. The typical range of $Pe_{d}$ in humans is such that advection dominates and there are no significant alterations to drug transport in the upper airways (see Fig.\ref{fig:Pe_vi}a). However, in the deep lungs, where there is no mucociliary advection, drug retention is enhanced at a larger $Pe_{d}$ (defined based on upper airway parameters) due to comparatively smaller diffusion (see Fig.\ref{fig:Pe_vi}a inset).

Drug molecule diffusivity ($D_{d}$) depends inversely on the drug molecule size and viscosity of the mucus (see Eq. \ref{eq:particle_scaling}). A smaller molecule and lower viscosity of the mucus would, therefore, inhibit drug retention in the deep lungs but would not significantly alter drug retention in the upper airways due to weak dependence on $Pe_{d}$. Controlling the size of the drug molecule and mucus property modification is therapeutically viable and can be a possible approach to enhance drug retention in the deep lungs without significantly impacting retention in the upper airways.

In pathophysiological conditions, if there is impaired mucociliary advection, then it may lead to significantly reduced $Pe_{d}$. Such a situation would promote drug retention in the upper airways since the time-scale for pure diffusive drug transport would be extremely long.

\subsection*{Effect of breathing time period on drug deposition and retention}
\label{sec:airway_strouhal}
Deposition of drug-laden aerosols and drug retention in the lungs also depends on the breathing time period $T_b$ through two parameters $-$ the airway Strouhal number $St_a$ (Eq. \ref{eq:aerosol_transport_final}) and the mucus Strouhal number $St_m$ (Eq. \ref{eq:part_tr_final}). $St_a$ is the ratio of the advective time scale of airflow to the breathing time period (see Eq. \ref{eq:aerosol_scaling}). A longer breathing time period leads to lower $St_a$. Keeping all other parameters the same, long breaths are ``deeper'' and lead to greater volume being inhaled. Consequently, the fraction of drug-laden aerosols deposited in the deep lungs are observed to increase as $St_a$ decreases (see Fig. \ref{fig:Pe_ae}c). Correspondingly, $\phi_{d}$ increases and shifts towards deeper airways (see Fig. \ref{fig:Pe_vi}b). It is seen that $\phi_{d}$ remains substantial in the deep lungs when $St_a\leq0.01$, but becomes negligible when $St_a\geq0.05$ (see \textit{Supplementary Material, Fig. S4} for more details). 

The breathing time period $T_b$ also impacts the mucus Strouhal number ($St_m$), which is the ratio of the mucociliary advection to breathing time scales (see Eq. \ref{eq:particle_scaling}). A longer breathing time period relative to the time scale of mucociliary advection leads to lower $St_m$, which implies greater advective clearance of the mucus in a breathing cycle. Thus, longer breaths inhibit drug retention (see Fig. \ref{fig:Pe_vi}c). This is particularly evident from the drug washout curve at $N=0$ (see Fig. \ref{fig:Pe_vi}c inset). However, lower drug retention is observed to remain limited to the upper airways and does not influence drug retention in the deep lungs (see \textit{Supplementary Material, Fig. S5} for more details). 

In summary, on the one hand, longer breath time period leads to deep lungs deposition of drugs, which is good. On the other hand, it also inhibits drug retention in the upper airways, which is bad. These conflicting outcomes can be resolved by noting that longer breaths do not affect drug retention in the deep lungs. Achieving deep lung deposition is more critical. Shorter breathing times or shallow breaths can reduce deep lungs deposition of the drug-laden aerosols. Similar observations have also been made in experimental investigations carried out by Mallik et al. \cite{mallik2020experimental}.

\subsection*{Effect of exposure time}

The impact of exposure duration ($\tau_{exp}$) is studied by varying the number of breathing cycles for which the lungs are assumed to be exposed to the drug-laden aerosols at the inlet of $N=0$ generation. It is observed that the aerosol deposition pattern within the lungs remains almost identical with increase in $\tau_{exp}$, but the magnitude of aerosol deposition ($S_d$) (and hence $\phi_{d}$) increases as $\tau_{exp}$ become longer (see \textit{Fig. S6}). This increases the washout time causing longer retention of drugs in the lungs. It is found that the increase in $S_d$ and $\phi_{d}$ with $\tau_{exp}$ is linear, as shown in Fig. \ref{fig:Pe_vi}d. This information can be used to estimate the exposure time required for achieving a required drug concentration in various regions of the lungs or to estimate the drug dose delivered to a particular region of the lung over a specific exposure time (see \textit{Supplementary Materials, Section IVC} for more details). 

\subsection*{Drug delivery efficacy}

Pulmonary drug delivery systems have a major drawback since majority of the inhaled aerosolized drugs get deposited in the mouth and the pharynx. Only about $5-12\%$ of the inhaled drugs reach the trachea for further inhalation into the respiratory tract \cite{sellers2013inhaled}. This often leads to prescription of larger drug doses in order to obtain the required health effects. Larger drug doses can, however, lead to side effects and the drug dose prescribed should, therefore, be minimized as much as possible. The present study helps in identifying plausible routes for enhancing the efficacy of drug delivery to the lungs and thereby, minimizing the inhaled drug dose.

\begin{table*}[!ht]
    \caption{Comparison of drug dose delivered to the deep lung for various aerosol sizes and breathing periods. The aerosols carry the drugs and are generated from inhalers. It is assumed that $10\%$ of the aerosols inhaled reach the trachea for further inhalation into the deep lung \cite{sellers2013inhaled}. Enhancement is calculated with respect to $3$ $\mu$m aerosols for 4s breathing period.}
    \begin{tabular}{m{2.5cm} m{2.5cm} m{2.5cm} m{2.5cm} m{2.5cm}}
    \hline 
    Inhaled Dose per puff ($\mu$g) & Aerosol Size ($\mu$) & Breathing Period ($s$) & Drug dose reaching deep lung per puff ($\mu$g) & Enhancement ($\%$)\\
    \hline
    \multirow{6}{*}{$100$} & 0.02 & \multirow{4}{*}{4} & $3.95$ & 41\\
    
    & 0.5 &  & $1.41$ & $-49.6$\\
    & 3 & & $2.8$ & n/a\\
    & 10  &  & $0.36$ & $-87.1$\\
    \cline{2-5}
    & \multirow{3}{*}{$3$}  & 2 & $0.52$ & $-81.4$\\
    & & 4 & $2.8$ & n/a\\
    & & 8 & $5.38$ & $92.14$\\
    & & 16 & $6.39$ & $128.21$\\
    \hline
    \end{tabular}
    \label{tab:drug_efficacy}
\end{table*}

For example, consider the delivery of salbutamol from a pressurised meter-dose inhalers ($100$ $\mu$g per puff) in an asthmatic child. It is estimated using the present analysis that only $2.8$ $\mu$g (out of $100$ $\mu$g) per puff of aerosolised salbutamol i.e. $2.8\%$ of the inhaled drugs reach the deep lung considering the size of the aerosolized drugs to be 3 $\mu$m (corresponding deposition fraction of $28\%$) and $10\%$ inhaled aerosols reaching the trachea (see Table \ref{tab:drug_efficacy}). Aerosols generated from inhalers are in the range of $1-5$ $\mu$m. The corresponding salbutamol concentration in blood is estimated to be $42.26$ ng/ml after 40 inhaler puffs assuming the total deposited drugs in the deep lung to remain available to blood circulation (see \textit{Supplementary Materials, Section IVC} for more details). 20-40 puffs, corresponding to $20-40$ ng/ml of salbutamol in blood, are usually required to reverse the effects of bronchoconstriction in children \cite{sellers2013inhaled}.  The present analysis can, thus, be used to obtain a close estimate of the physiologically measured drug concentration. This can be used to gauge the efficacy of drug delivery for various combination of the pertinent parameters.

The present analysis shows that a plausible way of increasing the efficacy of drug delivery to the deep lung is by controlling the size of the inhaled aerosols generated using inhalers/nebulizers. Drug delivery to the deep lung is observed to be reduced significantly if the corresponding aerosol size is larger than $5$ $\mu$m or smaller than $1$ $\mu$m (see Table \ref{tab:drug_efficacy}). Aerosols larger than $5$ $\mu$m deposit mainly in the upper airways due to impaction, while those smaller than $1$ $\mu$m mostly remain suspended and are exhaled out resulting in lower deposition in the lung \cite{sung2007nanoparticles}. However, drug delivery to the deep lung increases substantially if $0.02$ $\mu$m aerosols are inhaled (see Table \ref{tab:drug_efficacy} and \textit{Fig. S2 in Supplementary Results for more details}) due to more efficient diffusional deposition of aerosols smaller than $0.1$ $\mu$m \cite{sung2007nanoparticles}. As such, drug delivery to the deep lung could be enhanced if such small aerosols are used. However, aerosols in this size range are impractical in the context of drug delivery systems because of the large energy requirement for generation of such aerosols \cite{sung2007nanoparticles}.

Controlling the time period of breathing while taking inhaler puffs (or using nebulizers) is another strategy which can be adopted to increase deep lung drug deposition. The present analysis shows that for longer breaths (see Table \ref{tab:drug_efficacy}) drug deposition increases significantly in the deep lung. Slow and deep breathing while inhaling the drugs can, as such, enhance the efficacy of deep lung drug deposition. This is the reason why it is recommended to breathe deeply and slowly while using inhalers/nebulizers \cite{shakshuki2017improving}.

\section*{Summary}
The present analysis uses a coupled aerosol (airway)-drug (mucus) flow model to determine deposition and retention of drug-laden aerosols in the lungs. It is observed that aerosols less than $10\text{ }\mu$m tend to deposit in the deep lungs (alveolar region) under normal breathing conditions. Deep lung deposition of aerosols have a non-monotonic dependence on aerosol sizes with maximum deposition for $0.02\text{ }\mu$m aerosols. Aerosol size and air flow rate, however, does not influence drug retention in the lungs. Longer breaths also promote deep lung aerosol deposition. However, longer breaths inhibit drug retention in the upper airways. Mucociliary clearance rate also controls the drug retention in the upper airways. In contrast, drug retention in the deep lungs depends only on diffusivity of the deposited drug molecules in mucus due to absence of mucociliary clearance. Thus, smaller drug molecules and lower mucus viscosity inhibits drug retention in the deep lungs by promoting quicker washout of the deposited drugs and vice-versa. 

Retention of the drugs in the lung is also observed to depend on the time for which the lungs are exposed to the drug-laden aerosols. The magnitude of aerosol (and hence, drug) deposition increases linearly with the exposure time with same qualitative nature. Larger deposition requires a longer washout period and hence, retention becomes enhanced with increase in exposure.   

Analysis further establishes that the efficacy of drug delivery to the deep lung can be enhanced by controlling the inhaled aerosol size and breathing time period. Drug delivery efficacy is observed to be maximum for aerosols in the size range of 1-5 $\mu$m. As such, aerosol generators like inhalers/nebulizers aim to produce aerosols in the above size range. Although larger efficacy are obtained for very fine aerosols ($<0.1$ $\mu$m) production of such aerosols are impractical in the context of pulmonary drug delivery. It is also observed that amount of drugs deposited in the deep lung increases by a factor of 2 when the breathing time period is doubled, with respect to normal breathing, suggesting breath control as a means to increase the efficacy of drug delivery to the deep lung.

\section*{Acknowledgements}
The authors gratefully acknowledge the grant provided by MHRD, Govt. of India under the SPARC programme (Project Code: P838).

\newpage
\centering
\section*{Appendix}
\justifying
\input{Appendix}

\bibliography{References}

\end{document}

%% file: Appendix.tex
\renewcommand{\thefigure}{A\arabic{figure}}
\setcounter{figure}{0}

\renewcommand{\thetable}{A\arabic{table}}
\setcounter{table}{0}

\renewcommand{\theequation}{A\arabic{equation}}
\setcounter{equation}{0}

\renewcommand{\thesection}{A\arabic{section}}
\setcounter{section}{0}

\section{\textbf{Idealization of the Lung Geometry}}

Table \ref{tab:parameters} summarises the magnitudes of various parameters used while approximating the lung model. Table \ref{tab:alveolated area fractions} lists the assumed fraction of airway area that is alveolated at each generation in the lung model.

\begin{table}[!h]
    \centering
    \caption{Parameters used in modelling the lung geometry}
    \begin{tabular}{c c c c}
    \hline
    $L_0$  &  0.12 m \cite{weibel1963morphometry}& $\alpha$ & 0.73\\
    $A_0$ &  $0.000317 \text{m}^2$ \cite{weibel1963morphometry}& $\beta$ & 0.71\\
    $R_0$ & $\sqrt{A_0/\pi}$ & $\zeta$ & 0.9\\
    $\delta_{0}$ & 10 $\mu$m \cite{karamaoun2018new}& $\varepsilon$ & 0.87\\
    $A_{m,0}$ & $2\pi R_0\delta_{0}$\\
    $V_{m,0}$ & $-5$ mm/min \cite{karamaoun2018new}& &\\
    \hline
    \end{tabular}
    \label{tab:parameters}
\end{table}

\begin{table}[!h]
    \centering
    \caption{Fractions of alveolated airways in different generations \cite{karthigathesis}}
    \begin{tabular}{c c}
    \hline
    Lung Generation ($N$) & Fraction of alveolated area ($\gamma$)\\
    \hline
    0-16 & 0\\
    17 & 0.0011\\
    18 & 0.0041\\
    19 & 0.0135\\
    20 & 0.0509\\
    21 & 0.1168\\
    22 & 0.2712\\
    23 & 0.5424\\
    \hline
    \end{tabular}
    \label{tab:alveolated area fractions}
\end{table}

\section{\textbf{Mathematical Model}}

\subsection{Aerosol transport}
\label{sec:droplet transport eqn}

The one-dimensional transport equation for aerosols at any location in the idealised lung geometry is expressed as
\begin{equation}
    \frac{\partial(A c_a)}{\partial t}+\frac{\partial(Qc_a)}{\partial x} = \frac{\partial }{\partial x}\Big(A D_a\frac{\partial c_a}{\partial x}\Big)-L_D c_a,
    \label{eq:aerosol_tr_1_app}
\end{equation}
where, $c_a$ represents the aerosol concentration, $Q$ represents the volume flow rate of air in breathing, and $D_a$ represents the diffusivity of aerosols in air. The coefficient $L_D$ accounts for the droplets deposited in the airway mucus. This equation is based on the \textit{trumpet} model proposed by Taulbee \& Yu \cite{taulbee1975theory} which has been later used by various authors to study different aspects of aerosol deposition in the lung \cite{devi2016designing,darquenne1994one,mitsakou2005eulerian}. The transport equation is formulated based on the assumption that the aerosols are monodispersed, do not undergo coagulation, and are decoupled from airflow in the lungs. It is also assumed that external forces (such as electrical and magnetic forces) do not have any influence on the aerosol dynamics. It is further assumed that there is no additional source of aerosols present within the lungs and the aerosols are either deposited in the airway mucus or washed out of the airways.

Eq. \ref{eq:aerosol_tr_1_app} is presented in terms of airway length ($x$), while the lung model adopted is in terms of lung generation number ($N$). As such, Eq. \ref{eq:aerosol_tr_1_app} needs to be converted to a more appropriate form in terms of $N$. This requires an additional mathematical relation (Eq. \ref{eq:length_to_Gen}) connecting airway length $x$ and the lung generation number ($N$) given by  

\begin{equation}
    H(N) = \dfrac{\partial N}{\partial x} = -\dfrac{1-\alpha}{L_0\alpha\text{ ln}(\alpha)\alpha^N}.
    \label{eq:length_to_Gen}
\end{equation}

Converting Eq.\ref{eq:aerosol_tr_1_app} using Eqs. \ref{eq:length_to_Gen} and $A_N=A_0(2\beta)^N$, we get

\begin{equation}
    A_0(2\beta)^N\frac{\partial c_a}{\partial t}=  H\frac{\partial }{\partial N}\Bigg[\Big(A_0(2\beta)^N D_aH\frac{\partial c_a}{\partial N}\Big)-\Big({Q_{max}q(t)c_a}\Big)\Bigg]-L_Dc_a,
    \label{eq:aerosol_tr_2}
\end{equation}
where, $q(t)$ represents the temporal sinusoidal function accounting for airflow variation during breathing such that $Q = Q_{max}q(t)$. Eq. \ref{eq:aerosol_tr_2} is reduced to its dimensionless form by multiplying and dividing Eq. \ref{eq:aerosol_tr_2} with $\bigg(\dfrac{L_0}{A_0 D_a}\bigg)$ and $\bigg(-\dfrac{\alpha \text{ ln}(\alpha)}{1-\alpha}\bigg)$, respectively, and using the following scaling parameters 
\begin{equation}
    \tau = \frac{t}{T_b}, 
    \phi_a = \frac{c_a}{c_{a,0}},
    T_a = \frac{L_0 A_0}{|Q_{max}|},
    St_a = \frac{T_a}{T_b},
    Pe_{a} = \frac{|Q_{max}|L_0}{A_0D_a},D_a=\dfrac{k_B T C_S}{3\pi\mu_{a}d_a},
    \label{eq:aerosol_scaling_app}
\end{equation}
where, $Pe_a$ and $St_a$ are the Peclet number for aerosols and Strouhal number for the airways, respectively. $\phi_a$ and $\tau$ denotes the dimensionless aerosol concentration and time, respectively, while the quantities $T_a$ and $T_b$ represents the convective airflow time-scale and breathing time-scale, respectively. The expression of $D_a$ is based on the Stokes-Einstein relation \cite{chakravarty2019aerosol}, where $C_s$ represents the Cunningham slip correction, $T$ represents the ambient temperature, $\mu_a$ denotes air viscosity, and $d_a$ denotes the aerosol diameter.  

The dimensionless equation, thus, obtained is used to analyse aerosol transport in the present study and is given by 
\begin{equation}
\begin{aligned}
    |Pe_{a}|St_a(2\alpha\beta)^N \frac{\partial( \phi_a)}{\partial \tau}=  \frac{\partial F_a}{\partial N}-L'_D\phi_a,
    \end{aligned}
    \label{eq:aerosol_transport_final_app}
\end{equation}
where, $L'_D$ represents the dimensionless form of aerosol deposition coefficient ($L_D$) and $F_a$ represents the total aerosol flux. These are expressed as follows - 
\begin{equation}
    L'_D=L_D\frac{L_0^2}{A_0D_a}\alpha^N
\end{equation}

\begin{equation}
    F_a = \Bigg[\Bigg(\Bigg(\frac{2\beta}{\alpha}\Bigg)^N \Bigg(\frac{1-\alpha}{\alpha (\text{ln} \alpha)}\Bigg)^2 \frac{\partial \phi_a}{\partial N}\Bigg)+\Big(|Pe_{a}|q(t)\Big(\frac{1-\alpha}{\alpha \text{ ln}(\alpha)}\Big)\phi_a\Big)\Bigg].
\end{equation}

\subsection{Aerosol deposition models}
\label{sec:droplet_deposition}

The major mechanisms of aerosol deposition in the lungs have been identified in the literature as diffusion, sedimentation and impaction of the aerosols in the airways, as well as diffusion and sedimentation of the aerosols in the alveoli \cite{hofmann2011modelling,devi2016designing}. Different empirical models have been used to estimate the different depositions. However, these models need to be converted into a more appropriate form for use in the present analysis. 

The probability of aerosol deposition in the airways by diffusion ($P_d$), sedimentation ($P_s$) and impaction ($P_i$) can be expressed following Yeh \& Schaum \cite{yeh1980models} as

\begin{equation}
    \dfrac{c_{a,0}-c_{a}}{c_{a,0}} = P_d+P_s+P_i-P_dP_s-P_dP_i-P_sP_i-P_dP_sP_i
    \label{eq:deposition_prob}
\end{equation}

Eq. \ref{eq:deposition_prob} can be re-written as

\begin{equation}
\begin{split}
   \dfrac{c_{a}}{c_{a,0}} & = (1-P_d)(1-P_s)(1-P_i)
    =(B_{d}e^{-k_{d}x})(B_{s}e^{-k_{s}x})(B_{i}e^{-k_{i}x})
    \\
    & \implies c_{a}
    =c_{a,0}\Bigg[ B_{d}B_{s}B_{i}e^{-(k_{d}+k_{s}+k_{i})x})\Bigg]
\end{split}
\label{eq:linearisation1}
\end{equation}
where, the terms $B_d$, $B_s$ and $B_i$ are the corresponding coefficients, and $k_d$, $k_s$ and $k_i$ are the corresponding constants in the exponential functions for different deposition mechanisms as proposed by Yeh \& Schaum \cite{yeh1980models}. Detailed expressions for the different deposition mechanisms can be found in the subsequent discussion. Taking the derivative of Eq. \ref{eq:linearisation1} with respect to $x$, we obtain - 

\begin{equation}
    \frac{d c_{a}}{d x}
    =-c_{a,0}\Bigg[(k_{d}+k_{s}+k_{i}) B_{d}B_{s}B_{i}e^{-(k_{d,n}+k_{s}+k_{i})x})\Bigg]
    = -(k_{d}+k_{s}+k_{i})c_{a}
    \label{eq:derivative1}
\end{equation}

Equation \ref{eq:derivative1} represents the droplet deposition flux in the airways. It is further converted to a dimensionally relevant form for use in the transport equation (Eq. \ref{eq:aerosol_tr_2}) as follows - 

\begin{equation}
    \frac{D (Ac_{a})}{D t}
\simeq Av\frac{dc_{a}}{dx}= -Av(k_{d}+k_{s}+k_{i})c_{a}
= L_Dc_{a}
\implies L_D = Av(k_{d}+k_{s}+k_{i})
    \label{eq:deposition_airways}
\end{equation}

The term $L_D$ represents the aerosol deposition coefficient which is determined using different empirical relations. The empirical relations are converted to a form relevant to Eq. \ref{eq:deposition_airways} and then reduced to their dimensionless forms for use in the final transport equation (Eq. \ref{eq:aerosol_transport_final_app}). These are discussed in the following sections for the various deposition mechanisms considered in this analysis. 

\subsubsection{Diffusional deposition in the airways}
\label{sec:diffusion_airways}
The probability of diffusional deposition of the aerosols in the airways can be expressed following Yeh \& Schaum \cite{yeh1980models} as 

\begin{equation}
\begin{split}
    P_d &= 1-0.819e^{-7.315Gx}-0.0976e^{-44.61Gx}-0.0325e^{-114Gx}\\
    &=1-B_{d,1}e^{-k_{d,1}x}-B_{d,2}e^{-k_{d,2}x}-B_{d,3}e^{-k_{d,3}x}
\end{split}
\label{eq:Diff_1}
\end{equation}
where, $G=\dfrac{D_{d}}{2R_N^2v_N}$.
The above equation can be simplified by expressing the coefficients in terms of effective magnitudes ($B_d$, $k_d$) as follows 

\begin{equation}
    P_d =1-B_{d}e^{-k_{d}x}
\label{eq:Diff_2}
\end{equation}
where, $B_d$ and $k_d$ are determined as

\begin{equation}
\begin{split}
    B_d &= B_{d,1}+B_{d,2}+B_{d,3}\\
    k_d &= k_{d,1}+k_{d,2}+k_{d,3}
\end{split}
\label{}
\end{equation}

\begin{figure}[!ht]
    \centering
    \includegraphics[scale=0.8]{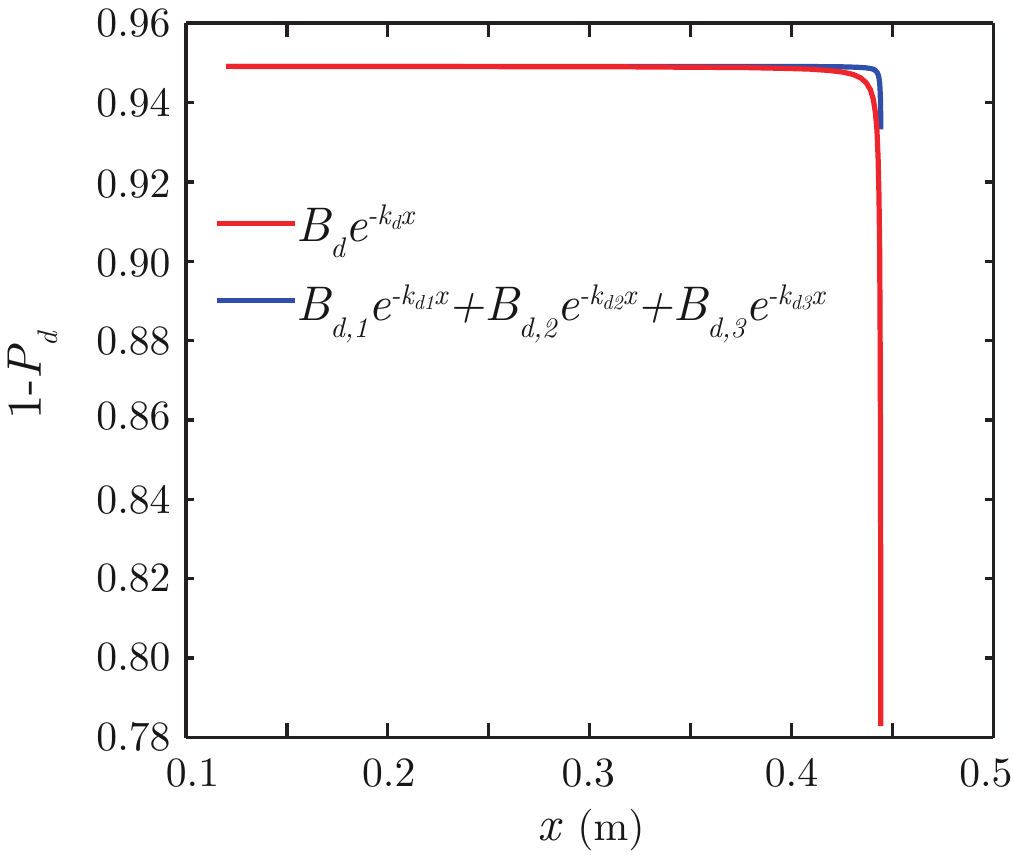}
    \caption{Comparison of the diffusional deposition probability using the simplified model ($B_d$, $k_d$) used in the present study and the model proposed by Yeh \& Schaum \cite{yeh1980models} for a aerosol diameter of $0.1\text{ }\mu$m.}
    \label{fig:comp_diff}
\end{figure}

It is estimated that this simplification does not have any significant influence on the calculation for diffusional deposition (see Fig. \ref{fig:comp_diff}). The simplified form is, as such, used in this analysis for calculation diffusional deposition in the airways.

Using Eqs. \ref{eq:Diff_1} and \ref{eq:Diff_2}, we obtain - 
\begin{equation}
k_{d,1} = 7.315\dfrac{D_{a}}{2R_N^2v_N},
k_{d,2} = 44.61\dfrac{D_{a}}{2R_N^2v_N},
k_{d,3} = 114\dfrac{D_{a}}{2R_N^2v_N}
\label{}
\end{equation}
and
\begin{equation}
k_d = k_{d,1}+ k_{d,2}+k_{d,3}= (7.315+44.61+114)\dfrac{D_{a}}{2R_N^2v_N}
\label{}
\end{equation}
where, $R_N$ and $v_N$ denotes the airway radius and airflow velocity of a particular lung generation, respectively. Using this, droplet deposition in the airways due to aerosol diffusion is estimated as

\begin{equation}
L_{D,d} = v_N A_{N,T} k_d = v_N \pi R_N^2 2^N (7.315+44.61+114)\dfrac{D_{a}}{2R_N^2v_N}
\label{eq:diff_dep}
\end{equation}

Conversion of Eq. \ref{eq:diff_dep} to its dimensionless form gives us the following expression for dimensionless diffusional deposition of the aerosols in the airways - 
\begin{equation}
L'_{D,d}=L_{D,d}\frac{L_0^2}{A_0D_{a}}\alpha^N = \Big(\dfrac{L_0}{R_0}\Big)^2 {(2\alpha)}^N (3.66+22.305+57)
\label{}
\end{equation}

\subsubsection{Sedimentation deposition in the airways}
\label{sec:sedimentation_airways}

The probability of deposition of the aerosols due to sedimentation in the airways is expressed following Yeh \& Schaum \cite{yeh1980models} as

\begin{equation}
    P_s = 1-exp\Bigg[-\Bigg(\frac{gC_s\rho_{a} d_{a}^2 \text{cos}(\psi_N)}{9\pi\mu_{air} R_N v_N}x\Bigg)\Bigg]    
    \label{}
\end{equation}
where, $\rho_{a}$, $g$ and $\psi$ represents droplet density, gravitational acceleration and airway orientation angle considering horizontal as $90^{\circ}$, respectively. Linearising the above equation using the approach followed in Eq.\ref{eq:linearisation1}, we obtain - 

\begin{equation}
    k_s = \frac{gC_s\rho_a d_a^2 \text{cos}(\psi_N)}{9\pi\mu_{air} R_N v_N}
    \label{}
\end{equation}

Aerosol deposition in the airways due to sedimentation can, then, be estimated as - 
\begin{equation}
\begin{aligned}
    L_{D,s} = v_N A_{N,T} k_s & = v_N(\pi R_N^2 2^N) \frac{gC_s\rho_a d_a^2 \text{cos}(\psi_N)}{9\pi\mu_{air} R_N v_N}\\
    & = \frac{1}{9}\frac{R_N g C_s\rho_a d_a^2 \text{cos}(\psi_N)}{\pi\mu_{air}}2^N
\end{aligned}
\label{}
\end{equation}

Conversion of the dimensional deposition ($L_{D,s}$) to its dimensionless form  gives us the following expression for dimensionless sedimentation deposition in the airways - 
\begin{equation}
\begin{aligned}
    L'_{D,s}=L_{D,s}\frac{L_0^2}{A_0D_{a}}\alpha^N = \frac{1}{3}\Big(\frac{L_0}{R_0}\Big)^2{(2\alpha\sqrt{\beta})}^NS_g \text{cos}(\psi_N)
\end{aligned}
\label{}
\end{equation}
where, $S_g$ is defined as the sedimentation parameter and expressed as

\begin{equation}
\begin{aligned}
    S_g = \frac{R_0\rho_a d_a^3 g}{k_BT}
\end{aligned}
\label{}
\end{equation}

\subsubsection{Impact deposition in the airways}
\label{sec:impact_airways}
The probability of deposition due to impaction of the aerosols in the airways is given by Yeh \& Schaum \cite{yeh1980models} as

\begin{equation}
P_i = 1-f_i(\theta,St)
\label{}
\end{equation}
where, $\theta$ denotes the branching angle of the airways and $St$ denotes the Stokes number ($=\dfrac{C_s\rho_a r_a^2v_N}{9\mu_{air}R_N}$). The function $f_i(\theta,St)$ is expressed as follows - 

\begin{equation}
\begin{split}
f_i(\theta,St) &= \frac{2}{\pi}\text{cos}^{-1}(\theta\cdot St)-\frac{1}{\pi}\text{sin}\Big[2\text{cos}^{-1}(\theta\cdot St)\Big] \text{ for } \theta\cdot St<1 \text{ (Inhalation)}\\
 &= 1 \text{ for } \theta\cdot St\geq1 \text{ (Exhalation)}
\end{split}
\end{equation}

The expression of $P_i$ is not in a form that can be directly linearised. As such, certain mathematical treatments need to be carried out in order to estimate the impact deposition. Loss of droplets in one generation of the lungs can be determined based on the droplet concentrations before and after the lung generation. Mathematically, this can be expressed as
\begin{equation}
\begin{split}
   \text{Loss in a generation} &= \frac{c_{bef}-c_{aft}}{c_{bef}} = 1-f_i(\theta,St)\\
\implies \frac{c_{aft}}{c_{bef}} &= f_i(\theta,St) 
\end{split}
\end{equation}

In terms of lung generations, the above expression can be re-written as
\begin{equation}
c_d = f_i^N(\theta,St)c_{a,0}
\label{}
\end{equation}

Differentiating with respect to generation number, we obtain - 
\begin{equation}
\dfrac{d c_a}{d N} = c_{a,0}\text{ln}(f_i^N(\theta,St))f_i^N(\theta,St) = \text{ln}(f_i^N(\theta,St))c_a
\label{}
\end{equation}

Converting the above derivative to a derivative in terms of $x$, we get - 
\begin{equation}
\begin{aligned}
\dfrac{d c_a}{d x} = -(-\text{ln}(f_i^N(\theta,St)))\frac{d N}{d x}c_a\\
\implies k_i = (-\text{ln}(f_i^N(\theta,St)))\dfrac{d N}{d x}
\end{aligned}
\label{}
\end{equation}

The impact deposition is estimated using the above expression as
\begin{equation}
L_{D,i} = v_N A_{N,T} k_i = v_N \pi R_N^2 2^N (-\text{ln}(f_i^N(\theta,St)))\dfrac{d N}{d x}
\label{}
\end{equation}

The dimensionless form of the impact deposition is obtained following a similar approach as in Sections \ref{sec:diffusion_airways} and \ref{sec:sedimentation_airways}. The dimensionless impact deposition, thus, obtained is expressed as
\begin{equation}
\begin{aligned}
    L'_{D,i}=L_{D,i}\frac{L_0^2}{A_0D_{a}}\alpha^N = |Pe_{a}|q(t)\text{ln}(f_i^N(\theta,St))\frac{(1-\alpha)}{\alpha \text{ ln}(\alpha)}
\end{aligned}
\label{}
\end{equation}

\subsubsection{Diffusional deposition in the alveoli}

Diffusional deposition of the aerosols in the alveoli is estimated using the following dimensionless expression - 

\begin{equation}
L'_{D,d,alv} = \gamma_N \eta_{d,alv}|Pe_{a}|q(t)\Big(\dfrac{1-\alpha}{-\alpha\text{ ln}(\alpha)}\Big)
\label{}
\end{equation}
where, $\gamma_N$ denotes the fraction of alveolated area in the corresponding generation (see Table \ref{tab:alveolated area fractions}) and $\eta_{d,alv}$ denotes the diffusional deposition efficiency in the alveoli. $\eta_{d,alv}$ is expressed as \cite{devi2016designing}

\begin{equation}
\eta_{d,alv} =1-\frac{6}{\pi^2}\sum\frac{1}{k^2}exp\Bigg[-\frac{4k^2tD_{a}}{d_{eq}^2}\Bigg] 
\label{}
\end{equation}

\subsubsection{Sedimentation deposition in the alveoli}

Deposition of the inhaled aerosols due to their sedimentation in the alveoli are estimated using the following dimensionless expression - 

\begin{equation}
L'_{D,s,alv} = \gamma_N \eta_{s,alv}|Pe_{a}|q(t)\Big(\dfrac{1-\alpha}{-\alpha\text{ ln}(\alpha)}\Big)
\label{}
\end{equation}
where, $\gamma_N$ and $\eta_{s,alv}$ denotes the fraction of alveolated area in the corresponding generation (see Table \ref{tab:alveolated area fractions}) and  sedimentation deposition efficiency in the alveoli, respectively. $\eta_{s,alv}$ is expressed as \cite{devi2016designing}

\begin{equation}
\eta_{s,alv} = {\Bigg[1+\text{min}\Big(\dfrac{d_s}{d_{eq}},1\Big)\Bigg]}^2 {\Bigg[1-0.5\text{min}\Big(\dfrac{d_s}{d_{eq}},1\Big)\Bigg]}^2-1
\label{}
\end{equation}

\subsection{Drug molecule transport in mucus}
\label{sec:virus transport eqn}

The corresponding 1D transport equation for the drugs deposited in the airway mucus is expressed as   

\begin{equation}
    \frac{\partial(A_{m} c_{d})}{\partial t}+\frac{\partial(Q_m c_{d})}{\partial x} = \frac{\partial }{\partial x}(A_{m} D_{d}\frac{\partial c_{d}}{\partial x})+\text{Source}
    \label{eq:virus_tr_1}
\end{equation}
where, $c_{d}$ denotes the drug concentration in the airway mucus, $Q_m$ represents the volume flow rate of mucociliary clearance and $D_{d}$ denotes the diffusivity of drug molecules in the mucus layer. 

The drug-laden aerosols deposited in the airway mucus serve as the only source of drugs in the lungs. The source term in Eq. \ref{eq:virus_tr_1} is, therefore, equivalent in magnitude to the deposition term in Eq. \ref{eq:aerosol_tr_1_app} ($L_{D}c_{a}$) times the drug load in aerosols ($\phi_l$). Mathematically, this is expressed as - 

\begin{equation}
    \text{Source} = L_{D}c_{a}\phi_l
    \label{eq:virus_source}
\end{equation}
where, $\phi_l$ is defined as the amount of drug molecules contained by the aerosols per unit amount of the aerosols. Equation \ref{eq:virus_tr_1} is converted to a form in terms of $N$ using Eqs. \ref{eq:length_to_Gen} and $A_{m}=A_{m,0}(2\sqrt{\beta}\zeta)^N$ in a similar manner as in Section \ref{sec:droplet transport eqn} as follows - 

\begin{equation}
\begin{aligned}
    A_{m,0}(2\zeta\sqrt{\beta})^N\frac{\partial c_{d}}{\partial t}= H\frac{\partial }{\partial N}\Bigg[\Big(A_{m,0}(2\zeta\sqrt{\beta})^N D_{d}H\frac{\partial c_{d}}{\partial N}\Big)-\Big(Q_{m,0}(2\varepsilon\zeta\sqrt{\beta})^N c_{d}\Big)\Bigg]+(\phi_l L_{D}c_{a})
\end{aligned}
    \label{eq:virus_tr_2}
\end{equation}

The above equation is further reduced by multiplying and dividing by $\Big(\dfrac{L_0}{A_{m,0} D_{d}}\Big)$ and $\Big(-\dfrac{\alpha \text{ ln}(\alpha)}{1-\alpha}\Big)$, respectively. The reduced equation is expressed as 

\begin{equation}
\begin{aligned}
    \frac{L_0 |V_{m,0}|}{D_{d}}(2\alpha\zeta\sqrt{\beta})^N\frac{T_m}{T_b}\frac{\partial c_{d}}{\partial t}= \frac{\partial }{\partial N}\Bigg[\Bigg(\Big(\frac{2\zeta\sqrt{\beta}}{\alpha}\Big)^N \Big(\frac{1-\alpha}{\alpha \text{ ln}(\alpha)}\Big)^2 \frac{\partial c_{d}}{\partial N}\Bigg)-\Bigg({\frac{L_0 |V_{m,0}|}{D_{d}}(2\varepsilon\zeta\sqrt{\beta})^N c_{d}\Bigg)}\Bigg]\\
    +\Bigg(\phi_l L'_{D}\frac{A_0D_{a}}{L_0^2\alpha^N}\phi_{a}c_{a,0}\frac{L_0^2 \alpha^N}{A_{m,0} D_{d}}\Bigg)
\end{aligned}
    \label{eq:virus_tr_3}
\end{equation}

The following parameters are utilised to achieve the dimensionless form of the virus transport equation in the airway mucus given by Eq. \ref{eq:virus_tr_final}.

\begin{equation}
    \tau = \frac{t}{T_b}, 
    \phi_{d} = \frac{c_{d}}{c_{d,0}},
    c_{d,0} = \phi_l c_{a,0} \dfrac{A_0}{A_{m,0}},
    T_m = \frac{L_0}{|V_{m,0}|},
    St_m = \frac{T_m}{T_b},
    Pe_{d} = \frac{|V_{m,0}|L_0}{D_{d}}, D_{d} = \dfrac{k_B T}{3\pi\mu_m d_{d}}
    \label{eq:virus_scaling}
\end{equation}

\begin{equation}
\begin{aligned}
    |Pe_{d}|(2\alpha\zeta\sqrt{\beta})^N St_m \frac{\partial \phi_{d}}{\partial \tau}= \frac{\partial F_{d}}{\partial N}+\Big(L'_{D}\frac{D_{a}}{D_{d}}\phi_{a}\Big)
    \end{aligned}
    \label{eq:virus_tr_final}
\end{equation}
where, $\phi_{d}$, $Pe_{d}$ and $St_{m}$ represents the dimensionless drug concentration, Peclet number for the drug molecules and Strouhal number for the mucus layer, respectively. $T_m$ denotes the time-scale for mucociliary transport. Drug diffusivity ($D_{d}$) is estimated using the Stokes-Einstein relation where $\mu_m$ represents mucus viscosity and $r_{d}$ represents size of the drug molecules. The term $F_d$ in Eq. \ref{eq:virus_tr_final} represents the total flux of the drug molecules and is expressed as

\begin{equation}
    F_{d} = \Bigg[\Bigg(\Big(\frac{2\zeta\sqrt{\beta}}{\alpha}\Big)^N \Big(\frac{1-\alpha}{\alpha \text{ ln}(\alpha)}\Big)^2 \frac{\partial \phi_{d}}{\partial N}\Bigg)-\Bigg(|Pe_{d}|(2\varepsilon\zeta\sqrt{\beta})^N \phi_{d})\Bigg)\Bigg]
\end{equation}

\subsection{Implementation of the model and validation}

The mathematical model discussed in Sections \ref{sec:droplet transport eqn}-\ref{sec:virus transport eqn} is implemented for computational analysis using MATLAB\textsuperscript{\tiny\textregistered}. The governing transport equations are discretised following the finite-difference technique with a first-order upwind and central-difference scheme used for the advective and diffusive terms, respectively. The temporal terms are discretised using explicit forward differencing. 

\begin{figure}[!ht]
    \centering
    \includegraphics[scale=0.75]{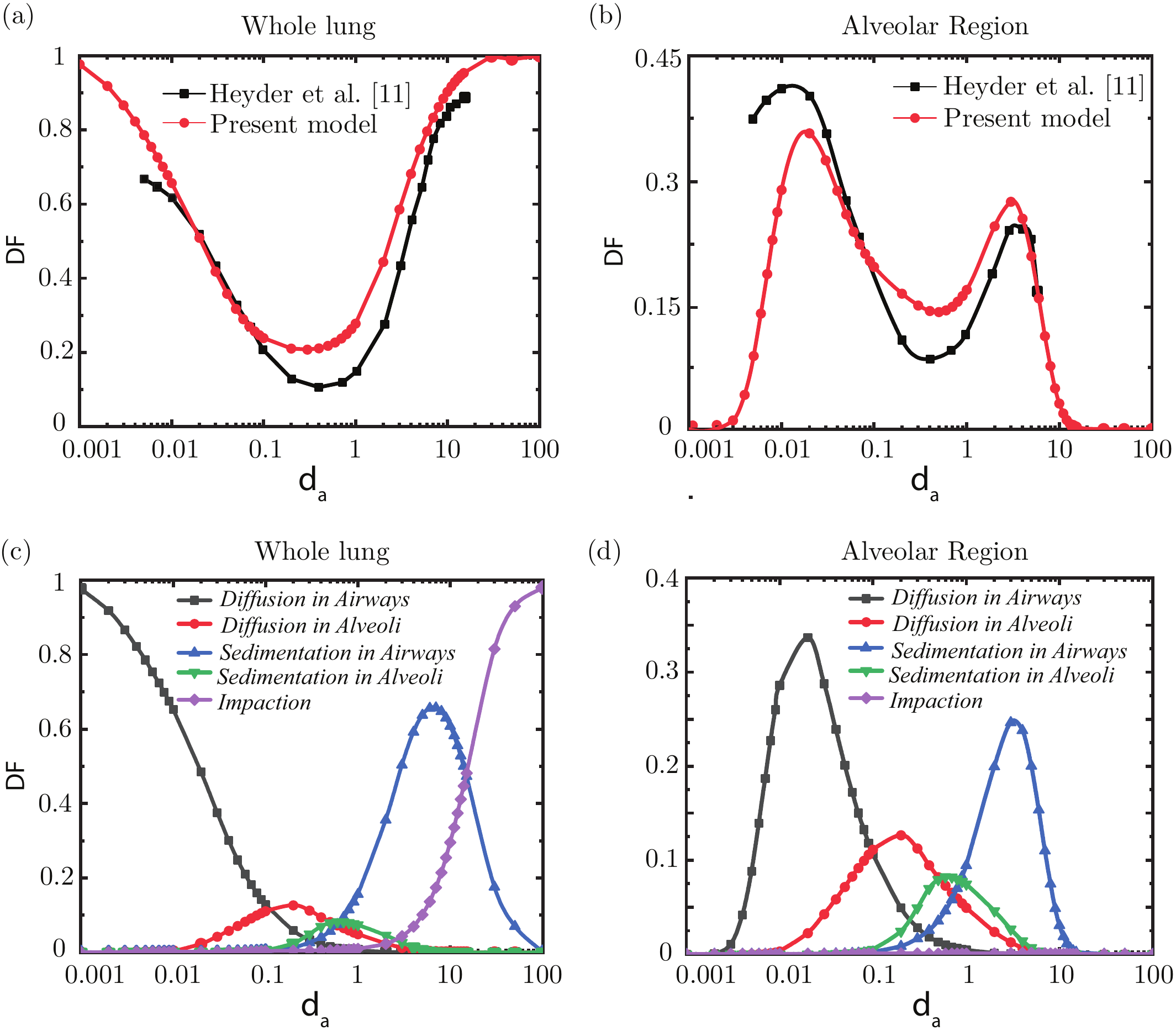}
    \caption{Comparison of the calculated deposition fraction ($DF$) of inhaled aerosols for (a) the whole lungs and (b) the alveolar region with the experimental results obtained by Heyder et al. \cite{heyder1986deposition} for different aerosol diameter ($d_{a}$), and comparison of the impact of different deposition mechanisms as a function of aerosol diameter in (c) the whole lung and (d) the alveolar region. }
    \label{fig:validation}
\end{figure}

The implemented mathematical model is validated with respect to aerosol deposition within the lungs. Aerosol depositions predicted using the computational model are compared with the experimental data of Heyder et al. \cite{heyder1986deposition} with respect to deposition in the whole lungs as well as deposition specifically in the alveolar region of the lungs. The results are shown in Figs. \ref{fig:validation}a-b. It can be observed that the computed aerosol deposition is in quite good agreement with the experimentally determined data. Figs. \ref{fig:validation}c-d represent the contribution of different deposition mechanisms considered in the present analysis in the whole lung as well as the alveolar region. The dominance of the different deposition mechanisms are similar to that observed from literature \cite{sznitman2013respiratory}.

\section{\textbf{Physiological basis for parameter selection}}

The magnitudes of different parameters used in the mathematical model are selected based on relevant physiological data. Physiological quantities pertinent to the lung model are tabulated in Table \ref{tab:parameters}. Other relevant physiological quantities are summarised in Table \ref{tab:physiological_data} below.

\begin{table}[!h]
    \centering
    \caption{Magnitudes of relevant physiological quantities considered in the study}
    \begin{tabular}{c c c c}
    \hline
    Quantity & Magnitude & Quantity & Magnitude\\
    \hline
    $d_a$  &  0.01-20 $\mu$m & $d_d$  & 0.01-0.1 $\mu$m\\
    $\mu_air$ & 0.000018 kg/ms & $\mu_m$ & 0.1 kg/ms\\
    $T$ & 300 K & $T_b$ & 4 s\\
    $Q_{max}$ & 0.0007925 $m^3/s$ & &\\
    \hline
    \end{tabular}
    \label{tab:physiological_data}
\end{table}

\section{\textbf{Supporting Results}}

\subsection{Effect of aerosol size on drug deposition in the deep lungs}

\begin{figure}[!ht]
\centering
\includegraphics[scale=0.75]{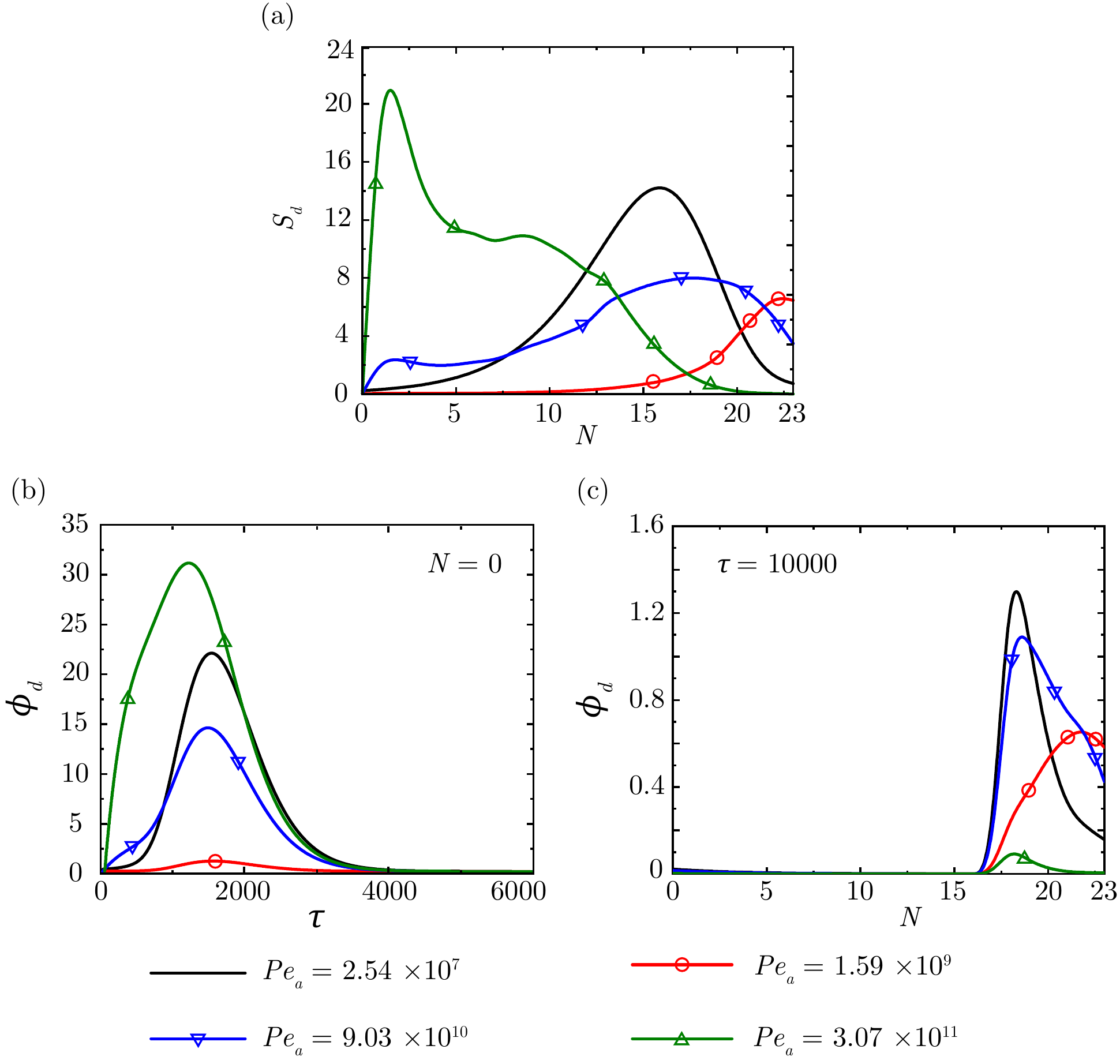}
\caption{(a) Aerosol deposition ($S_d=L'_{D}\phi_a$) within the lungs for different $Pe_{a}$ (b) Temporal change in drug concentration ($\phi_d$) at $N=0$ for different $Pe_{a}$ (c) Drug concentration within the lungs at $\tau=10000$ for different $Pe_{a}$. The results are shown for $St_a = 0.0095$, $Pe_{d} = 4.56 \times 10^{7}$, $St_m = 359.7122$, $\tau_{exp}=5$.}
\label{fig:Pe_ae_supp}
\end{figure}

\begin{figure}[!ht]
\centering
\includegraphics[scale=0.75]{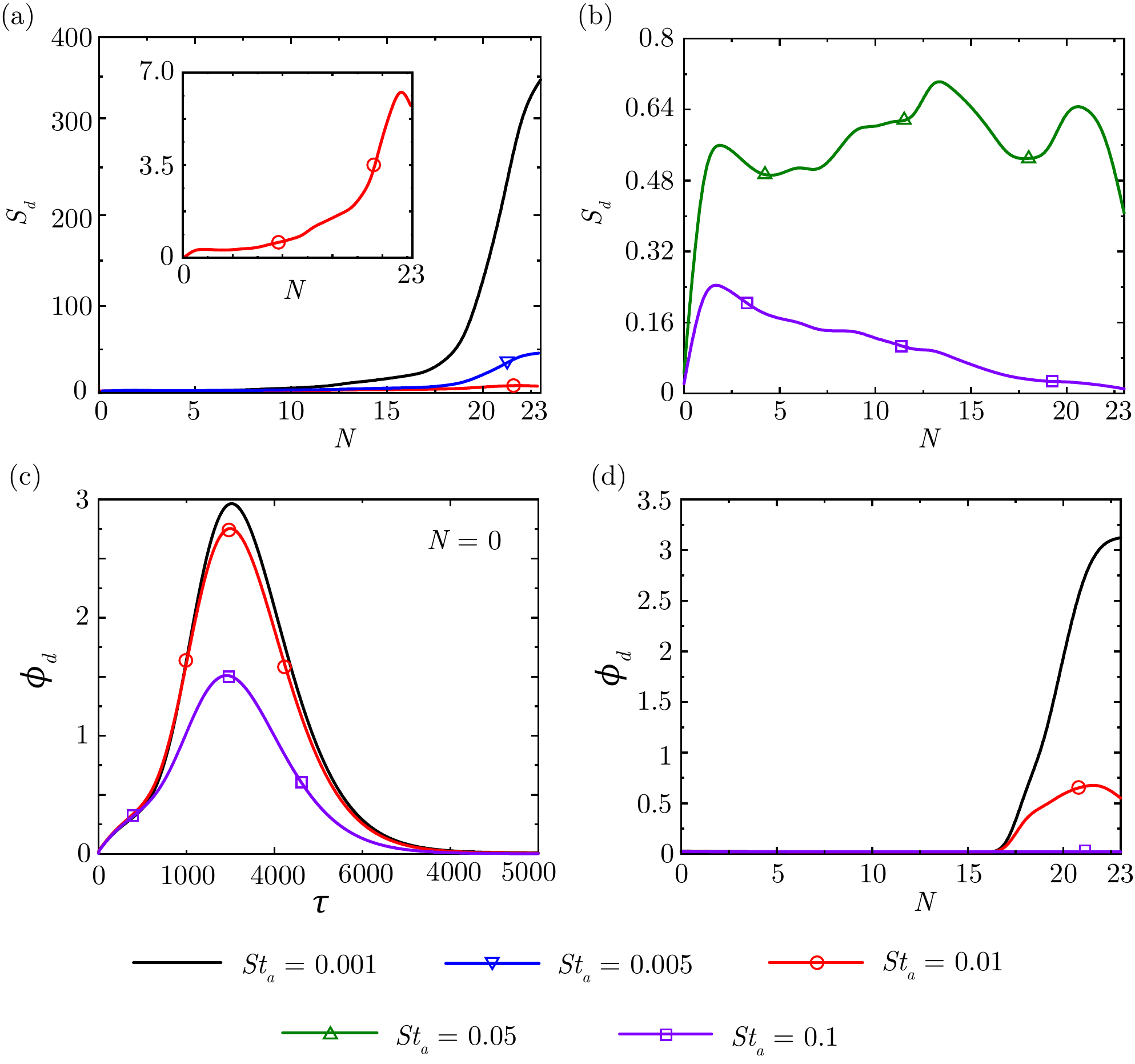}
\caption{a-b) Aerosol deposition ($S_d=L'_{D}\phi_a$) within the lungs for different $St_a$ (c) Temporal change in drug concentration ($\phi_d$) at $N=0$ for different $St_a$ (d) Drug concentration ($\phi_d$) within the lung for different $St_a$ at $\tau = 10000$. The results are shown for $Pe_{a} = 2.85 \times 10^{10}$, $Pe_{d} = 4.56 \times 10^{7}$, $St_m = 359.7122$, $\tau_{exp}=5$.}
\label{fig:St_a_supp}
\end{figure}

\begin{figure}[!ht]
\centering
\includegraphics[scale=0.75]{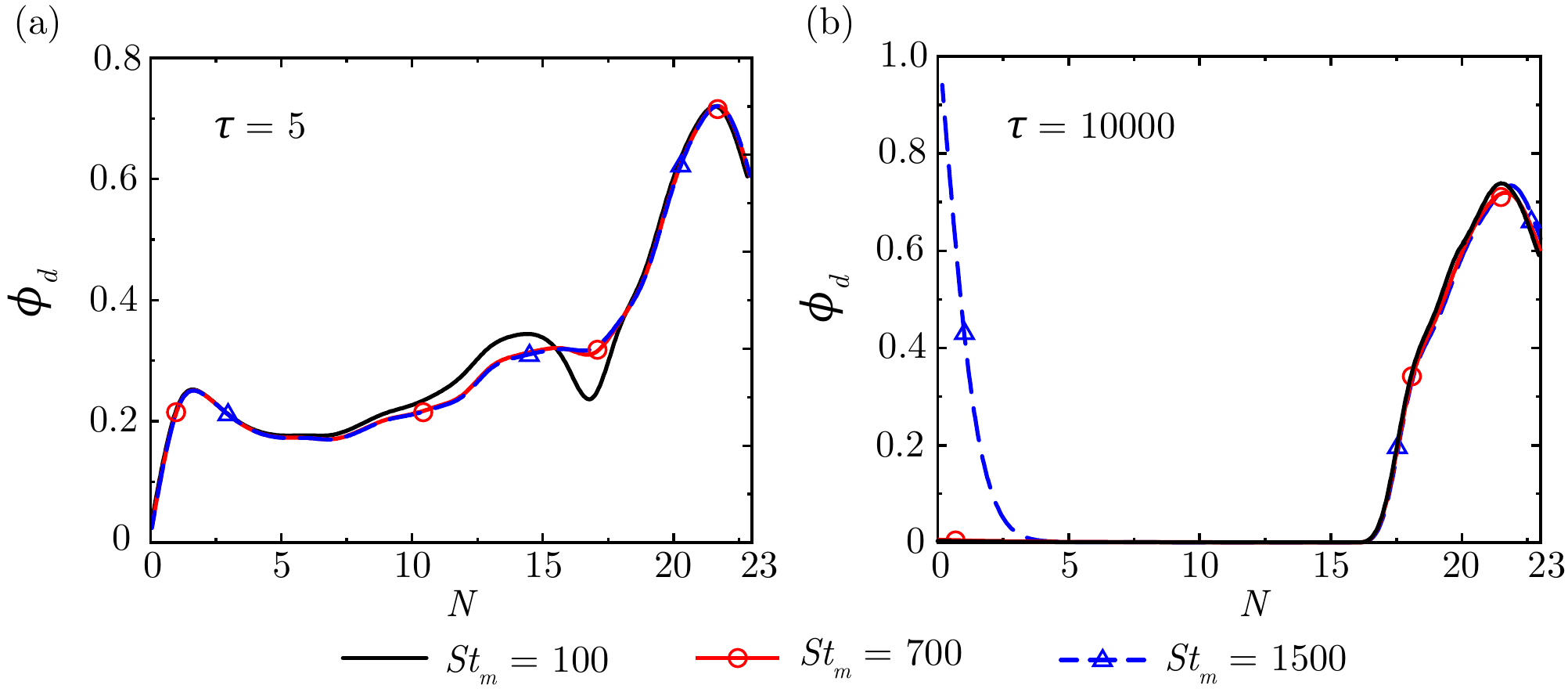}
\caption{Drug concentration ($\phi_v$) within the lungs for various $St_m$ at (a) the end of aerosol exposure ($\tau = 5$) and (b) at $\tau = 10000$. The results are shown for $Pe_{a} = 2.85 \times 10^{10}$, $Pe_{d} = 4.56 \times 10^{7}$, $St_a = 0.0095$, $\tau_{exp}=5$. }
\label{fig:St_m_supp}
\end{figure}

Fig. \ref{fig:Pe_ae_supp}a shows the variation in aerosol deposition ($S_d$) within the lung with change in $Pe_{a}$. A larger volume of aerosols are able to reach the deeper generations of the lung with increase in $Pe_{a}$ leading to larger aerosol deposition. This reverses when $Pe_{a}$ is increased beyond $1.59 \times 10^{9}$ and at $Pe_{a} = 3.07 \times 10^{11}$, most of the aerosols are observed to get deposited in the first few generations and almost no deposition in the deep lung (beyond $N=18$). Similar observations have been made in previous investigations as well \cite{choi2007mathematical}. The reason is due to a larger contribution of impact deposition of the aerosols in the earlier generations at such high $Pe_{a}$. The corresponding drug concentrations ($\phi_d$) are shown in \textit{Figs. 3a-b} in the main manuscript.

Fig. \ref{fig:Pe_ae_supp}b shows the temporal variation in $\phi_d$ at $N=0$ of the lung. It can be observed that washout of the drugs from the lung does not undergo any significant temporal change with variation in $Pe_{a}$. However, the initial location of drug deposition within the lung is observed to have an important impact on its washout. Drugs deposited before $N=18$ gets washed out quickly due to the stronger muco-ciliary clearance. Drugs deposited beyond $N=18$, however, gets transported much slowly due to the weak diffusive transport of the drug molecules in mucus in that region. Drugs deposited at very large $Pe_{a}$ ($\sim 3.07 \times 10^{11}$) are, therefore, washed out of the lungs relatively quickly since majority of the deposition takes place before $N=18$. At lower $Pe_{a}$, however, a substantial amount of the deposited drugs continue to persist in the deep lung (beyond $N=18$) even though muco-ciliary clearance washes out the drugs from the upper generations. The retention of drugs in the deep lung is evident from the distribution of $\phi_{d}$ within the lung at $\tau = 10000$ in Fig. \ref{fig:Pe_ae_supp}c.

\subsection{Effect of breathing time period on drug deposition and retention}
Figs. \ref{fig:St_a_supp}a-b highlights the change in aerosol deposition within the lung with variation in $St_a$. It can be observed that the magnitude of aerosol deposition in the mucus decreases and the deposition also tends to shift towards the upper airways with increase in $St_a$. This happens since the amount of aerosols being inhaled reduces with increase in $St_a$. The progression of $\phi_a$ front into the lung, therefore, decreases which, in turn, results in the aforementioned change in aerosol deposition pattern. The corresponding change in drug concentration ($\phi_d$) is shown in \textit{Fig. 4a} in the main manuscript.

Any change in $St_a$, however, do not affect the mucuociliary transport in the lung or drug diffusivity in the mucus. Drug washout from the lung, therefore, remains unaffected when $St_a$ is changing, as shown from the temporal change of drug concentration at $N=0$ in Fig. \ref{fig:St_a_supp}c. Persistence of drugs in the deep lung is, as such, observed for the situations where deep lung deposition of drugs occur i.e. $St_a\leq0.01$, as shown in Fig. \ref{fig:St_a_supp}d.

Figure \ref{fig:St_m_supp}a represents $\phi_d$ at the end of aerosol exposure for various $St_m$. It can be observed that there is no significant difference between $\phi_d$ when $St_m$ remains large. It is only when $St_m$ becomes $\sim 100$ that deviations become apparent enough. The reason for these deviations is the much faster mucus clearance at low $St_m$ which is able to transport the deposited drugs away from the initial deposition location even before the deposition is complete. 

Fig. \ref{fig:St_m_supp}b shows $\phi_d$ at $\tau = 10000$ for various $St_m$. It can be observed that there is a considerable difference between $\phi_d$ in the upper airways as a result of the varying rate of mucociliary transport. However, it is not observed to influence washout of the drugs from the deep lung in any manner. Breathing and muco-ciliary transport are, hence, observed to have no significant influence on drug washout from the deep lung.

\subsection{Effect of exposure time on drug deposition and retention}

\begin{figure}[!ht]
\centering
\includegraphics[scale=0.75]{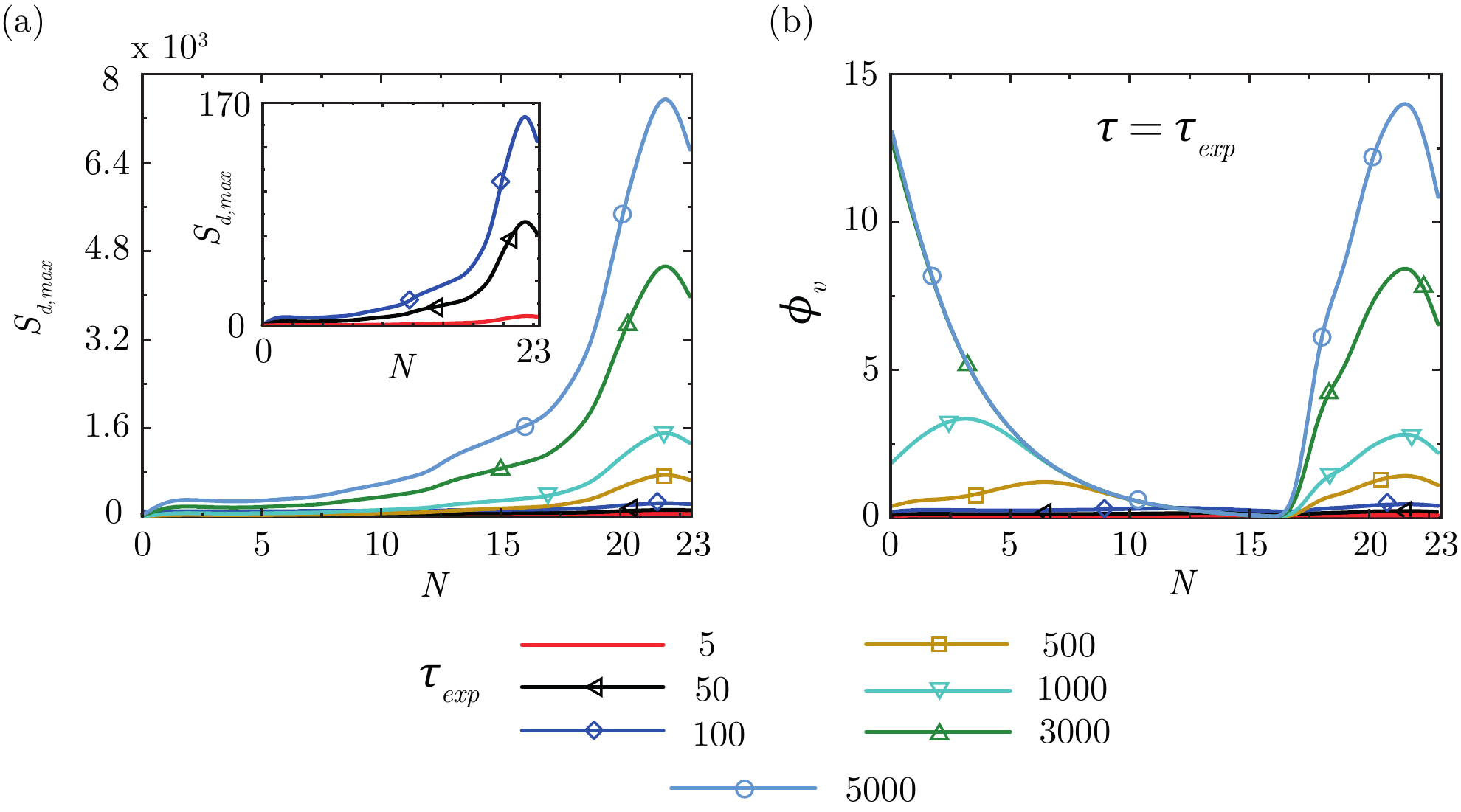}
\caption{(a) Total aerosol deposition ($S_d=L'_{D}\phi_a$) within the lung for different $\tau_{exp}$. Deposition for $\tau_{exp} = 5-100$ is additionally shown as inset to ensure proper readability (b) Drug concentration ($\phi_d$) within the lungs for different $\tau_{exp}$ at the end of exposure i.e. at $\tau = \tau_{exp}$. The results are shown for $Pe_{a} = 2.85 \times 10^{10}$, $Pe_{d} = 4.56 \times 10^{7}$, $St_a = 0.0095$, $St_m=359.7122$.}
\label{fig:exp_dur_supp}
\end{figure}

Figure \ref{fig:exp_dur_supp}a shows the total amount of aerosols deposited in the airway mucus for various $\tau_{exp}$ considered in this analysis. It can be observed that while the deposition pattern within the lungs remain almost identical, the magnitude of deposition increases as $\tau_{exp}$ become longer. The increase in deposition with exposure time is linear (see \textit{Fig. 4d} in the main manuscript). This observation can be used to estimate the dose of drugs that is delivered to a specific lung region over a particular period of time.

For example, pressurised meter-dose inhalers deliver $100$ $\mu$g of salbutamol per puff and it usually takes 20-40 puffs to reverse the effects of bronchoconstriction \cite{sellers2013inhaled}. Majority of the inhaled aerosolised drugs are deposited in the mouth and the pharynx, and only approximately $10\%$ of the inhaled aerosolised drugs reach the trachea for further inhalation. A maximum of $28\%$ of the aerosols that reach the trachea has been observed to reach the deep lung (corresponding aerosol size of 3 $\mu$m). Considering the above parameters, it is estimated that only $2.8$ $\mu$g per puff i.e. $2.8\%$ of inhaled drugs is able to reach the deep lung under normal breathing conditions.  Thus, for 40 puffs of inhaler, the total drug dose reaching the deep lung would be $112$ $\mu$g. Assuming the entire drug dose deposited in the deep lung to be passed on to the blood circulation, the estimated drug concentration in blood would be $42.26$ ng/ml considering the blood volume in children to be $2650$ ml. Salbutamol concentration of $20-40$ ng/ml in blood is considered adequate for reversing bronchoconstriction in children \cite{sellers2013inhaled}. Detailed calculation for the above estimation is as follows - 

\begin{equation*}
\begin{split}
    \textit{Dose per puff} &= 100 \mu\text{g}\\
    \textit{Dose per puff reaching the trachea} &= \textit{Dose per puff} \times \textit{Fraction of inhaled drugs reaching the trachea}\\
    &= 100 \mu\text{g} \times 10\%\\
    \textit{Dose per puff reaching deep lung} &= \textit{Dose per puff reaching trachea} \times\\
    &\textit{  Fraction of inhaled drugs at trachea reaching deep lung}\\
    &= 100 \mu\text{g} \times 10\% \times 28\%\\
    &= 2.8 \mu\text{g}\\
    \textit{Total dose reaching deep lung} &= \textit{Dose per puff reaching deep lung} \times \textit{Number of puffs}\\
    &= 2.8 \mu\text{g} \times 40\\
    &= 112 \mu\text{g}\\
    \textit{Drug concentration in blood} &= \textit{Total dose reaching deep lung}/\textit{Total blood volume}\\
    &= 112 \mu\text{g}/2650 \text{ ml}\\
    &= 42.26 \text{ ng/ml}
\end{split}
\end{equation*}

Similar calculations can be carried out for other combination of the pertinent parameters. The computational model can, thus, be utilised to estimate drug deposition and also to suggest ways to improve the drug delivery to the deep lung. Although the magnitude of drug deposition and drug concentration after a certain exposure duration can be determined by such extrapolations, it needs to be noted that this method is not a substitute for detailed simulations. Detailed simulations are still needed for drug retention calculation. The deposition characteristics can also change with variation in any one of the relevant parameters. Also, this knowledge does not provide information about the fraction of the inhaled aerosols that are deposited in the deep lungs. These informations can only be obtained from detailed simulations.

Figure \ref{fig:exp_dur_supp}b shows the drug concentration within the lungs for various $\tau_{exp}$ at the end of respective exposures. As expected, the drug concentration also increases due to larger aerosol deposition. In the upper airways ($N<18$), mucuociliary transport clearance occurs simultaneously with aerosol deposition and as such, the effective drug concentration is the resultant of drug transport due to the deposition and clearance mechanisms. While drug concentration increases in these generations due to higher aerosol deposition, continuous mucus transport clears the drugs from these generations towards the $0^{th}$ generation and as a consequence, drug accumulates in the first few generations (leading to much higher $\phi_{d}$) before being washed out of the lung.